\documentclass[aps,prd, nofootinbib, floatfix, notitlepage,twocolumn]{revtex4-1}

\usepackage{amsmath,amsfonts,amssymb}
\usepackage{mathrsfs}
\usepackage{graphicx}
\usepackage[english]{babel} 
\usepackage{hyperref} 
\usepackage{color}
\usepackage{xfrac}
\usepackage{bm}
\usepackage{physics}
\usepackage{makecell}
\usepackage{multirow}
\usepackage{layouts}
\usepackage[utf8]{inputenc}
\usepackage[T1]{fontenc}
\usepackage{ifthen}
\usepackage{verbatim}
\newcommand\mnras{MNRAS}
\newcommand\apjs{Astrophys. J. Supp.}
\newcommand\apjl{Astrophys. J. Lett.}

\newcommand{\G}{\ensuremath{\mathcal{G}}}
\newcommand{\D}{\ensuremath{\mathrm{d}}}
\newcommand{\mean}[1]{\ensuremath{\langle #1 \rangle}}
\newcommand{\meanM}[1]{\ensuremath{\left\langle #1 \right\rangle}}
\newcommand{\be}{\begin{equation}}
\newcommand{\ee}{\end{equation}}
\newcommand{\bes}{\begin{equation*}}
\newcommand{\ees}{\end{equation*}}

\newcommand{\resub}[1]{#1}

\newcommand{\dve}[4][1]{
	\ifthenelse{\equal{#1}{1}}{\left.\dv{#2}{#3}\right\vert_{#4}}{\left.\dv[#1]{#2}{#3}\right\vert_{#4}}}
\newcommand{\pdve}[4][1]{
	\ifthenelse{\equal{#1}{1}}{\left.\pdv{#2}{#3}\right\vert_{#4}}{\left.\pdv[#1]{#2}{#3}\right\vert_{#4}}}

\begin{document}

\title{Ultralight axions and the kinetic Sunyaev-Zel'dovich effect}
\author{Gerrit S. Farren$^{1,2}$}
\email{gfarren@haverford.edu}
\email{gsf29@cam.ac.uk}
\author{Daniel Grin$^2$}
\author{Andrew H. Jaffe$^3$}
\author{Ren\'ee Hlo\v zek$^{4,5}$}
\author{David J. E. Marsh$^{6}$}

\affiliation{$^{1}$Department of Applied Mathematics and Theoretical Physics, University of Cambridge, Centre for Mathematical Sciences, Wilberforce Road, Cambridge CB3 0WA, United Kingdom}
\affiliation{$^2$Department of Physics and Astronomy, Haverford College, 370 Lancaster Avenue, Haverford, Pennsylvania 19041, United States}
\affiliation{$^3$Astrophysics Group \& Imperial Centre for Inference and Cosmology, Department of Physics,
	Imperial College London, Blackett Laboratory, Prince Consort Road, London SW7 2AZ, United Kingdom}
\affiliation{$^{4}$Dunlap Institute for Astronomy and Astrophysics, University of Toronto, 50 St George Street, Toronto, Ontario, M5S 3H4, Canada}
\affiliation{$^{5}$David A. Dunlap Department of Astronomy and Astrophysics, University of Toronto, 50 St George Street, Toronto, Ontario, M5S 3H4, Canada}
\affiliation{$^{6}$
Theoretical Particle Physics and Cosmology, King's College London, Strand, London, WC2R 2LS}

\begin{abstract}
Measurements of secondary cosmic microwave background (CMB) anisotropies, such as the Sunyaev-Zel'dovich (SZ) effect, will enable new tests of neutrino and dark sector properties. The \textit{kinetic} SZ (kSZ) effect is produced by cosmological flows, probing structure growth. Ultralight axions (ULAs) are a well-motivated dark-matter candidate. Here, the impact of ULA dark matter (with mass $10^{-27}$ to $10^{-23}~{\rm eV}$) on kSZ observables is determined, applying new analytic expressions for pairwise cluster velocities and Ostriker-Vishniac signatures in structure-suppressing models. For the future CMB Stage 4 and ongoing Dark Energy Spectroscopic Instrument galaxy surveys, the kSZ effect (along with primary anisotropies) will probe ULA fractions $\eta_a = \Omega_{\rm{axion}}/\Omega_{\rm DM}$ as low as $\sim 5\%$ if $m_{a}\simeq 10^{-27}~{\rm eV}$ (at 95\% C.L.), with sensitivity extending up to $m_{a}\simeq 10^{-25}~{\rm eV}$. If reionization and the primary CMB can be adequately modeled, Ostriker-Vishniac measurements could probe values $\eta_{a}\simeq 10^{-3}$ if $10^{-27}~{\rm eV}\lesssim m_{a}\lesssim 10^{-24}~{\rm eV}$, or $\eta_{a}\simeq 1$ if $m_{a}\simeq 10^{-22}~{\rm eV}$, within the fuzzy dark matter window.

\end{abstract}

\date{\today}

\maketitle

\section{Introduction}

A standard cosmological model has been established, using measurements of cosmic microwave background (CMB) anisotropies \cite{Crites:2014prc,Aghanim:2018eyx,Akrami:2018vks,Aiola:2020azj}, determinations of cosmic acceleration from Type Ia supernovae \cite{Abbott:2018wog}, and the clustering/lensing of distant galaxies \cite{Abbott:2018wzc}. In this $\Lambda$ cold-dark matter ($\Lambda$CDM) model, the cosmic energy budget consists of baryons, nonrelativistic dark matter (DM), neutrinos, and ``dark energy" (DE), with relic-density parameters of $\Omega_{\rm b}h^{2}=0.0224\pm 0.0001 $, $\Omega_\textrm{c}h^{2}=0.1200 \pm 0.0012$, and $\Omega_{\rm DE}=0.6847 \pm 0.0073 $ \cite{Akrami:2018vks}.

The standard model (SM) of particle physics does not contain compelling DM or DE candidates, signaling (along with the hierarchy problem \cite{Susskind:1978ms}, the strong $\mathcal{CP}$ problem \cite{Peccei:1977hh}, and neutrino mass \cite{Gouvea:2016shl}) that new physics is needed. Future observations will include cosmic-variance limited measurements of CMB polarization using the Simons Observatory (SO) \cite{Galitzki:2018avl}, CMB Stage 4 (CMB-S4)  \citep{Abazajian:2019eic}, and extensive maps of large-scale structure (LSS) by the Vera C. Rubin Observatory \cite{Bechtol:2019acd}, the Dark Energy Spectroscopic Instrument (DESI) \cite{Aghamousa:2016zmz}, the Nancy Grace Roman Space Telescope \cite{2015arXiv150303757S}, and the \textit{Euclid} satellite \cite{Amendola:2012ys}. 

These efforts will test dark-sector physics, probing neutrino masses  \cite{Galitzki:2018avl,Abazajian:2019eic,Abazajian:2019eic}, the number of light relics \cite{Galitzki:2018avl,Abazajian:2019eic,Abazajian:2019eic}, and the DE equation-of-state parameter  \cite{Amendola:2012ys,2015arXiv150303757S}, as well as physical properties of DM \cite{Li:2018zdm}. The CMB's sensitivity to new physics will depend on \textit{secondary} anisotropies \cite{1980MNRAS.190..413S,Dodelson:1993xz,1995ApJ...442....1P,Ma:2001xr,Dore:2003ex,Santos:2003jb}, such as CMB lensing and the Sunyaev-Zel'dovich (SZ) effect, caused by the Compton scattering of CMB photons by free electrons \cite{Sunyaev:1970er,1980MNRAS.190..413S,Birkinshaw:1998qp}. 

The SZ effect induces a CMB intensity change proportional to $(v_{e}/c)\tau$, where $\tau$ is the scattering optical-depth and $v_{e}$ is the electron velocity. The SZ contribution from thermal electrons is known as the \textit{thermal} SZ (tSZ) effect \cite{Sunyaev:1970er,Sunyaev:1980vz}, while the bulk-flow contribution is known as the \textit{kinetic} (kSZ) effect \cite{1980MNRAS.190..413S,1995ApJ...442....1P}. The tSZ effect is measured using its nonthermal spectrum \cite{AtacamaCosmologyTelescope:2020wtv}. In contrast, the kSZ effect has a thermal spectrum and responds to real-time structure growth  \cite{1995ApJ...455..419P,Alonso:2016jpy}, as electron peculiar velocities scale as $v_{\rm pec, e}\propto \dot{\delta}
$ by the continuity equation, where $\delta$ is a fractional overdensity and dots denote time derivatives \cite{1995ApJ...455..419P,Ferreira:1998id,Zhang:2008pwa,Park:2015jea,Ma:2017ybt,Yasini:2018rrl}. The rms kSZ imprint on the CMB is $\sim10~{\mu}{\rm K}$ and suppressed as $v/c$ (compared to the tSZ effect),  making detection challenging. 

Nonetheless, the kSZ effect due to bulk flows  \cite{Hand:2011ye,Hand:2012ui,Hernandez-Monteagudo:2015cfa,Ma:2017ybt,Calafut:2021wkx} has been detected, using the cross-correlation of Atacama Cosmology Telescope (ACT) CMB maps with Sloan Digital Sky Survey (SDSS) LRG and CMASS galaxy data \cite{Hand:2011ye,Hand:2012ui}, as well as other data, such as \textit{Planck} and South Pole Telescope (SPT) maps of the CMB, and the Baryon Oscillation Spectroscopic Survey \cite{Hernandez-Monteagudo:2015cfa,Ma:2017ybt,Li:2017uin}. The kSZ signature of mildly nonlinear fluctuations [known as the Ostriker-Vishniac (OV) effect] could test models of cosmic reionization \cite{1986ApJ...306L..51O,1987ApJ...322..597V,Jaffe:1997ye,Scannapieco:2000ut,Ma:2001xr,Castro:2002df,Castro:2002dfE,Zahn:2005fn,Iliev:2006zz,Diego:2007mv,Lee:2009wa,Hernandez-Monteagudo:2009ydj,Calabrese:2014gwa}.

Future kSZ measurements could probe neutrino masses down to $\simeq33~{\rm meV}$ \cite{Mueller:2014dba}, $\sim5\%$-level changes to the DE equation of state \cite{DeDeo:2005yr,Bhattacharya:2007sk,Mueller:2014nsa}, and  deviations from general relativity \cite{Mueller:2014nsa,Bianchini:2015iaa}. We determine the response of kSZ observables to ultralight axions (ULAs), hypothetical particles that could contribute to the dark sector \cite{Arvanitaki:2009fg,Marsh:2015xka,Stott:2017hvl,Grin:2019mub}.

\begin{figure}
    \centering
    \includegraphics[trim={0.5cm 0.15cm 1.0cm 0.15cm},clip, width=0.9\columnwidth]{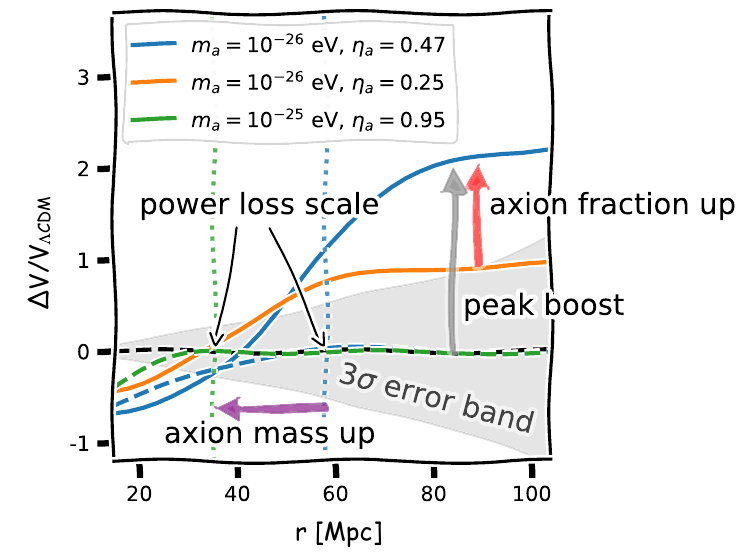}
    \caption{Impact of axions on mean pairwise velocities of galaxy clusters as a function of separation $r$, relative to $\Lambda$CDM predictions. Curves are obtained as described in Secs. \ref{sec:mpv_theory}-\ref{sec:vel}. Velocities for unbiased tracers (\textbf{dashed lines}) are suppressed below characteristic scales (\textbf{dotted lines}), where density fluctuations drop to 90\% of $\Lambda$CDM values. Due to structure suppression, fixed tracers are higher significance peaks in the density, making them more biased (\textbf{peak boost} behavior), and thus enhanced relative to $\Lambda$CDM on large scales (\textbf{solid lines}). The level of enhancement is dependent on the axion abundance $\eta_a$.}
    \label{fig:dmp}
\end{figure}
ULAs (with $10^{-33}~{\rm eV}\lesssim m_{a}\lesssim 10^{-10}~{\rm eV}$) are ubiquitous in string-inspired scenarios, e.g.~as Kaluza-Klein modes of fields in extra dimensions \cite{Conlon:2006tq,Svrcek:2006yi,Arvanitaki:2009fg,Cicoli:2012sz,Marsh:2015xka,Mehta:2021pwf}, and behave as ``fuzzy" DM (FDM) \cite{Hu:2000ke}. If $m_{a}\gtrsim 10^{-27}~{\rm eV}$, ULAs begin to dilute as matter (with density $\rho \propto a^{-3}$, for scale factor $a$) before matter-radiation equality. There could be an ``axiverse" of ULAs of many masses, with one solving the strong-$\mathcal{CP}$ problem \cite{Peccei:1977hh,Weinberg:1977ma,Wilczek:1977pj,Kim:1979if,Shifman:1979if,Dine:1981rt,Zhitnitsky:1980tq,Svrcek:2006yi,Arvanitaki:2009fg,Stott:2017hvl}. 

Via SM interactions, ULAs could be detected using experiments and astronomical observations \cite{stadnik:2014prd,abel:2017prx,Sigl:2018fba,Fedderke:2019ajk,Pogosian:2019jbt,Gruppuso:2020kfy,Bianchini:2020osu,Ejlli:2020yhk,Namikawa:2020ffr,Budker:2013hfa,Graham:2017ivz,Payez:2014xsa,Ivanov:2018byi,Ouellet:2018beu,Bogorad:2019pbu,Niemeyer:2019aqm}, though we focus on gravitational effects \cite{Arvanitaki:2009fg,Grin:2019mub,Hui:2021tkt,Niemeyer:2019aqm}. ULAs suppress clustering on galactic scales due to their large de Broglie wavelengths \cite{Khlopov:1985jw,Nambu:1989kh,Hu:2000ke,Suarez:2011yf,2012PhRvD..86h3535P,Hlozek:2014lca,Urena-Lopez:2015gur}. For masses $m_{a}\gtrsim 10^{-22}~{\rm eV}$, ULAs mitigate challenges to $\Lambda$CDM, such as Milky-Way satellite populations \cite{Martinez-Medina:2014hka,Schive:2015kza,Veltmaat:2016rxo,Mocz:2017wlg,Schive:2017biq,Veltmaat:2018dfz,Mocz:2018ium,Mocz:2019pyf,Mocz:2019uyd,Safarzadeh:2019sre,Nadler:2020prv} and galaxy cores \cite{Schive:2014hza,Martinez-Medina:2015jra,Du:2016aik,Bernal:2017oih,Du:2018zrg,DeMartino:2018zkx,deMartino:2018krg}. ULAs would alter the black-hole mass spectrum and gravitational-wave signatures \cite{Arvanitaki:2010sy,Pani:2012vp,Khmelnitsky:2013lxt,Ikeda:2018nhb,Stott:2018opm,Kitajima:2018zco,Baumann:2018vus,DeMartino:2017qsa,Porayko:2018sfa}. ULAs may even Bose condense \cite{Guth:2014hsa,Hertzberg:2016tal,Tsujikawa:2021rpg}. For values $m_{a}\leq 10^{-23}~{\rm eV}$, data allow a $\sim 1-5\%$ ULA contribution to DM \cite{Frieman:1995pm,Amendola:2005ad,Hlozek:2014lca,Hlozek:2017zzf,Poulin:2018dzj}. ULA-like particles could resolve cosmological tensions \cite{Riess:2019cxk,Poulin:2018cxd,Lin:2019qug,Agrawal:2019lmo,Smith:2019ihp,Hill:2020osr,Smith:2020rxx,Blum:2021oxj,Allali:2021azp,Lague:2021frh}, such as the $\sim 5\sigma$ tension between CMB and supernovae inferences of the Hubble constant $H_0$.

 CMB primary temperature anisotropies have been used to impose the limit $\Omega_{a}h^{2}\leq 6 \times 10^{-3}$ at the
$95\%$ C.L. \cite{Hlozek:2014lca} if $10^{-32}~{\rm eV}\lesssim m_{a}\lesssim 10^{-25.5}~{\rm eV}$, while polarization and CMB lensing data require $\Omega_{a}h^{2}\leq 3\times 10^{-3}$, with considerable sensitivity extending to $m_{a}\simeq 10^{-24}~{\rm eV}$ \cite{Hlozek:2017zzf}. Using lensing, future efforts like SO and CMB-S4 will probe values as low as $\Omega_{a}h^{2}\leq 2\times 10^{-4}$ \cite{Hlozek:2016lzm,Galitzki:2018avl,Abazajian:2019eic}, with improvements from galaxy lensing \cite{Marsh:2011bf} and intensity mapping \cite{Bauer:2020zsj}. The power of CMB lensing motivates us to determine how ULAs alter the kSZ effect.

We derive and evaluate the OV power spectrum in the presence of  structure-suppressing species (focused on ULAs, but with applications to neutrinos and ark energy). We find that ULA fractions of $\sim 10^{-3}$ might be probed using future OV measurements. So far, kSZ detections have been made by taking the difference between CMB temperature measurements in the directions of galaxy clusters \cite{Ferreira:1998id,Zhang:2008pwa,Bhattacharya:2007sk,Hand:2012ui,Sugiyama:2015dsa,Sugiyama:2016rue,Smith:2018bpn}, probing their pairwise velocities. Clusters (with masses $M\sim 10^{14} M_{\odot}$) are the heaviest collapsed objects, and their mass function responds to ULAs \cite{Diehl:2021gna}. We apply the halo model  \cite{1977ApJS...34..425D,Bardeen:1985tr,Sheth:1999su,Sheth:2000fe,Sheth:2000ff,Sheth:2000ii,Sheth:2001dp,Jenkins:2000bv,Cooray:2002dia,Ludlow:2016ifl,Bose:2015mga,Marsh:2016vgj} to explicitly derive (to our knowledge for the first time in the literature) expressions for cluster pairwise velocities in structure-suppressing scenarios, which
differ from those in Refs. \cite{Bhattacharya:2007sk,Mueller:2014dba,Mueller:2014nsa}, with more physical behavior at small scales.
 
We use 
\textsc{AxionCAMB}\footnote{\textsc{AxionCAMB} \cite{Hlozek:2014lca}, available at
\url{http://github.com/dgrin1/axionCAMB}, is a modified version of the Boltzmann code \textsc{CAMB} \cite{lewis:2000}. The version of \textsc{AxionCAMB} used here is found at \url{http://github.com/gerrfarr/axionCAMB}. The code used for kSZ predictions is available at \url{https://github.com/gerrfarr/Axion-kSZ-source}.} \cite{Hlozek:2014lca,Hlozek:2016lzm,Hlozek:2017zzf} to obtain power spectra and perturbation growth rates. We compute pairwise velocities, which are suppressed at small scales. Our results are summarized by Fig.~\ref{fig:dmp}. Compared to $\Lambda$CDM, cluster galaxies are 
rarer, more biased, peaks in density, \emph{enhancing} velocities at large separations (as noted in Refs. \cite{Bauer:2020zsj,Lague:2021frh}).

The effect can be large compared to typical peculiar velocitis; the residual is as large as $200~{\rm km}~{\rm s}^{-1}$ at comoving separations $r=50~{\rm Mpc}~h^{-1}$ for $m_a=5\times 10^{-26}~{\rm eV}$ or about $\sim 1.5$ times the $\Lambda$CDM velocity at $r\geq 50~{\rm Mpc}~h^{-1}$. We perform a sensitivity forecast, finding that CMB/LSS data at S4 \cite{Abazajian:2019eic} and DESI \cite{Aghamousa:2016zmz} sensitivity levels will probe ULA fractions of $\Omega_{\rm a}/\Omega_{\rm DM}\sim 10^{-2}$ for $m_{a}\simeq 10^{-27}~{\rm eV}$ (with comparable sensitivity up to $m_{a}\simeq 10^{-25}~{\rm eV}$). 

We begin in Sec.~\ref{sec:struc_form} by summarizing cosmological aspects of ULAs. We continue in Sec.~\ref{sec:signature} by deriving kSZ observables in ULA scenarios, beginning with the OV effect, and continuing with pairwise halo velocity signatures. In Sec.~\ref{sec:detectable}, we obtain numerical predictions as well as a Fisher-matrix forecast for the sensitivity of kSZ measurements to ULAs. We conclude in Sec.~\ref{sec:conc}. Expressions for the OV power spectrum are derived in Appendix
\ref{app:OV_derivation}. Halo-model derivations are found in Appendix \ref{app:mpv_derivation}, while some numerical integration techniques/parameter degeneracies are discussed in Appendixes \ref{app:num} and \ref{sec:degen}, respectively.

\section{ULA structure formation}
\label{sec:struc_form}

ULAs with cosmologically relevant densities have extremely high occupation numbers and may be modeled as a classical wave (see Refs. \cite{Marsh:2015xka,Kuss:2021gig} and references therein). The ULA energy density is roughly constant at early times and then transitions to DM-like dilution with the cosmic expansion. Gradient energy in the scalar field prevents localization of ULAs on length scales smaller than their de Broglie wavelength $\lambda_{\rm dB}=1/(m_{a}v)$, leading to the suppression of growth in cosmological structure for comoving wave numbers $k>2\pi/(a\lambda_{\rm dB})$ \cite{Marsh:2015xka}. 

 The fractional temperature difference induced by inverse Compton scattering of CMB photons (the kSZ effect) off ionized material in the intergalactic medium is given by the integral along the line of sight \cite{1986ApJ...306L..51O}
\begin{equation}
\frac{\Delta T}{T}=- \int n_e \sigma_T e^{-\tau}\left[\bm v( \chi\hat{\bm r}, a) \cdot \hat{\bm{r}}\right] a\ \D \chi, \label{eq:ddtkz}
\end{equation}
where $\chi$ is the comoving distance along the line of sight, $n_e$ is the electron density, $\sigma_T$ is the Thomson cross section, $\tau$ is the optical depth to $\chi$ and $\bm v(\bm{\chi}, a)$ is the bulk electron velocity field. The unit vector $\hat{\bm r}$ points along the line of sight. 

On the other hand, it can be shown from the continuity equation (e.g. Refs. \cite{Ma:1995ey,Ma:2001xr}) that on subhorizon scales the bulk electron velocity with Fourier wave vector ${\bm k}$ (and magnitude $k=|{\bm k}|$) is given to linear order by \cite{Jaffe:1997ye}
\begin{equation}\label{eq:velocity_kspace}
\tilde{\bm v}(\bm k, t)= \frac{i a}{k^2} H(a)\frac{\G(k, a)}{\G_0(k)} \dv{\ln \G}{\ln a} \bm k \delta_0(\bm k),
\end{equation}
where the growth factor $\mathcal{G}(k,a)$ describes the time dependence of density perturbations,
\begin{equation}
\mathcal{G}(k,a)\equiv\frac{\delta(\mathbf{k},a)}{\delta(\mathbf{k},a=1)}
\end{equation} and the $0$ subscript stands for the present day ($a=1$).

As a result, the ULA-induced contribution to cosmic structure formation will modify observations affected by the kSZ effect. To assess this effect quantitatively, we must first determine the evolution of linear perturbations in ULA models. We begin with a summary of the changes to linear cosmological perturbation theory induced by ULAs, following closely the treatment in Ref.  \cite{Hlozek:2014lca}.

The background ULA field $\phi_{0}$ obeys the Klein-Gordon (KG) equation in an expanding homogeneous Friedmann-Robertson-Walker spacetime, which is
\begin{equation}
\phi_{0}''+2\mathcal{H}\phi_{0}'+m_a^2a^2\phi_0=0,\label{eq:kghom}
\end{equation}where $m_a$ is the ULA mass in natural units, $a$ is the cosmological scale factor, $\mathcal{H}=a'/a$ is the conformal Hubble parameter, and $'$ denotes a derivative with respect to conformal time $\eta$, defined by $d\eta=dt/a$. ULAs make a contribution 
\begin{equation}
\rho_{a}=\frac{\phi_{0}'^{2}}{2a^{2}}+\frac{m_a^{2}\phi_{0}^{2}}{2a^{2}}\label{eq:rhosfhom}
\end{equation}
to the total energy density and 
\begin{equation}
P_{a}=\frac{\phi_{0}'^{2}}{2a^{2}}-\frac{m_a^{2}\phi_{0}^{2}}{2a^{2}}\label{eq:Psfhom}
\end{equation} to the total pressure, working in the quadratic approximation to the full ULA potential [$V(\phi)\propto (1-\cos{\phi/f_{a}})]\simeq \phi^{2}/(2f_{a}^{2})$, which is valid through most of the parameter space of observational interest \cite{Hlozek:2014lca}.\footnote{See Refs. \cite{Schive:2017biq,Arvanitaki:2019rax} for a discussion of interesting phenomena in halo cores and linear-theory mode growth in the strongly anharmonic portion of the potential.}

Early on, $\phi_{0}$ rolls slowly with equation-of-state parameter (EOS) $w_{a}\equiv P_{a}/\rho_{a}\simeq -1 $. Once the Hubble parameter $H$ has fallen sufficiently for the condition  $3H\ll m_{a} $ to be satisfied, the field coherently oscillates with a period $\Delta t \sim 1/m_{a}$ and so the cycle-averaged energy dilutes as matter. In other words, $\langle \rho_{a }\rangle\propto a^{-3}$ and $\langle w_{a}\rangle\simeq 0$, where the brackets $\langle\rangle $ denote a cycle average \cite{Cookmeyer:2019rna}. The transition between these regimes occurs when $a=a_{\rm osc}$, defined by $m=3H(a_{\rm osc})$. 

If this transition occurs prior to matter-radiation equality (after which most modes responsible for galaxy formation enter the horizon), that is, if $a_{\rm osc}\leq a_{\rm eq}$ (matter-radiation equality), we may think of ULAs as `DM-like', because they begin to dilute as DM prior to the horizon entry of the modes relevant for large-scale structure formation. 

On the other hand, if this transition occurs after equality (if $a_{\rm osc}\geq a_{\rm eq}$), standard galaxy formation is altered if ULAs are considered as a component of dark \emph{matter}. In this case, we can think of ULAs as `DE-like'. The boundary between these two regimes occurs for a value $m_{a}\sim 10^{-27}~{\rm eV}$. When using a halo-model approach with $m_{a}\lesssim 10^{-27}~{\rm eV}$, there are subtle complications that arise in determining if  (and for which scales) ULAs should be treated as a clustering component of the cosmological density field \cite{Marsh:2013ywa,Hlozek:2014lca,Bauer:2020zsj}. Here, we restrict our attention to DM-like ULAs, and defer these lower-$m_{a}$ complications for future investigation.

Perturbations to the ULA fluid (denoted $ \phi_1$) obey the perturbed version of Eq.~(\ref{eq:kghom}), with additional terms due to metric perturbations, which are sourced by ULAs and SM fields through the Einstein equations. For a ULA field fluctuation with Fourier wave vector $\vec{k}$,  ULA contributions to the metric are determined by their energy density perturbation $\delta \rho_{a}$, pressure perturbation $\delta P_{a}$, and momentum flux $u_{a}$,

\begin{align}
\delta \rho_{a}=&a^{-2}\phi_{0}'\phi_{1}'+m_{a}^{2}\phi_{0}\phi_{1}-a^{2}{\dot{\phi}}_{0}^{2}A,\label{eq:drho} \\
\delta P_{a}=&a^{-2}\phi_{0}'\phi_{1}'-m_{a}^{2}\phi_{0}\phi_{1}-a^{2}{\dot{\phi}}_{0}^{2}A,\label{eq:dp}\\
u_{a}-(1+w_a)B=&k \frac{\phi_{0}'\phi_{1}}{\rho_{a}a^{2}}\label{eq:dmom},
\end{align} 
where $A$ is the scalar metric perturbation and $B$ is the longitudinal vector perturbation (in any chosen gauge). The first term in both of Eqs.~(\ref{eq:drho}) and (\ref{eq:dp}) is the perturbative expansion of the canonical kinetic term for small field fluctuation, while the second term comes from perturbations to a quadratic potential. Equation~(\ref{eq:dmom}) expresses the velocity perturbation in terms of conformal-time derivative of the background field and fluctuations $\phi_{1}$.

For our purposes, these perturbations are conveniently (and exactly) described using the generalized dark matter (GDM) equations of motion (EOM) \cite{Hu:1998kj}, with Fourier-space continuity and Euler equations that may be derived directly from the perturbed KG equation. They are given in synchronous gauge by 
\begin{align}
\delta_{a}'=&-ku_{a}-(1+w_{a})h_{L}'/2-3\mathcal{H}\left(1-w_{a}\right)\delta_{a}\nonumber \\&-9\mathcal{H}^{2}\left(1-c_{\rm ad}^{2}\right)u_{a}/k,\label{eom:gexact_cont}\\
u_{a}'=&2\mathcal{H}u_{a}+k\delta_{a}-\frac{w_{a}'}{k\left(1+w_{a}\right)}u_{a}\label{eom:gexact_euler}
,\end{align} where $k$ is the Fourier wave mode number of the fractional ULA density perturbation $\delta_{a}=\delta\rho_{a}/\rho_{a}$ and its corresponding value of $u_{a}$. 

The term proportional to $u_{a}$ in the continuity equation, Eq.~(\ref{eom:gexact_cont}), is present due to mass flux out of infinitesimal volumes. The remaining terms in Eq.~(\ref{eom:gexact_cont}) are gauge-dependent terms of relevance for superhorizon modes. The synchronous gauge time-time metric perturbation is $h_{L}$, following the conventions of Ref. \cite{Ma:1995ey}, which we use throughout this discussion. The term proportional to $h_{L}'$ is present due to redshift in the presence of a local gravitational field.

The sole term on the left-hand side and last term on the right-hand side of Eq.~(\ref{eom:gexact_euler}) arise from terms of the form $dp/d\eta$ in the standard Euler momentum-conservation equation. The first term on the right-hand side of Eq.~(\ref{eom:gexact_euler}) corresponds to the redshifting of nonrelativistic momentum in an expanding Friedmann-Robertson-Walker background. The second term on the right-hand side of Eq.~(\ref{eom:gexact_euler}) represents the impact of pressure gradients on fluid velocities.

In addition to the EOS parameter $w_{a}$, fluid perturbation evolution is governed by the adiabatic sound speed
\begin{equation}
c_{\rm ad}^{2}\equiv \frac{P_{a}'}{\rho_{a}'}= w-\frac{w_{a}'}{3\mathcal{H}\left(1+w_{a}\right)}
.\end{equation}
In terms of GDM variables, the ULA contributions to the $00$ and trace of the $ii$ Einstein equations are 
\begin{align}
\delta \rho_{a}=&\rho_{a}\delta_{a},\label{eq:source_a}\\
\delta P_{a}=&\rho_{a}\left[\delta_{a}+3\mathcal{H}\left(1-c_{\rm ad}^{2}\right)\frac{u_{a}}{k\left(1+w_{a}\right)}\right].\label{eq:source_b}
\end{align}
The GDM EOMs [Eqs.~(\ref{eom:gexact_cont})-(\ref{eom:gexact_euler})] are an \textit{exact} restatement of the perturbed KG equation. They become prohibitively expensive to solve with sufficient accuracy for cosmological observables when $a\gg a_{\rm osc}$, however because coherent oscillations occur much faster than the Hubble expansion, resulting in rapid oscillation of Einstein-equation terms that couple background pressure oscillations, metric fluctuations, and field perturbations \cite{Fan:2016rda}.

To ease this difficulty, we follow past work \cite{Hu:2000ke,Hwang:2009js,Hlozek:2014lca,Suarez:2015fga,Urena-Lopez:2015gur,Cembranos:2015oya,Fan:2016rda,Desjacques:2017fmf,LinaresCedeno:2018oso,Poulin:2018dzj,Cookmeyer:2019rna} and use an effective fluid approximation (EFA). This approximation is obtained by taking a cycle average of perturbed fluid variables and restating the perturbed KG equation into a gauge in which the cycle average $\langle u_{a}\rangle=0$. Recasting the perturbed KG equation in terms of perturbed fluid variables [applying Eqs.~(\ref{eq:source_a}) and (\ref{eq:source_b}) and transforming back into synchronous gauge], the following continuity and Euler equations are obtained \cite{Hwang:2009js,Hlozek:2014lca,Poulin:2018dzj,Cookmeyer:2019rna}\footnote{In the limit that $w_{a}=w_{a}'=c_{\rm ad}^{2}=0$ for the exact equations and $c_{s}^{2}=1$ for the EFA, the two sets of EOMs agree, but we stress that while Eqs.~(\ref{eom:gexact_cont}) and (\ref{eom:gexact_euler}) are exact, Eqs.~(\ref{eq:efa_cont}) and (\ref{eq:efa_euler}) apply (and are used) deep in the rapidly oscillating regime.}:
\begin{align}
\delta_{a}'=&-ku_{a}-\frac{h_{L}'}{2}-3\mathcal{H}c_{s}^{2}\delta_{a}-9\mathcal{H}^{2}c_{s}^{2}u_{a}/k,\label{eq:efa_cont}\\
u_{a}'=&-\mathcal{H}u_{a}+c_{s}^{2}k\delta_{a}+3c_{s}^{2}\mathcal{H}u_{a}.\label{eq:efa_euler}
\end{align}The EFA is essentially an implementation of the Wentzel–Kramers–Brillouin (WKB) approximation, averaging over the ULA field's rapid oscillations and encoding the structure growth suppression of the model with a scale-dependent \textit{effective} sound speed \cite{Hwang:2009js,Poulin:2018dzj}:
\begin{equation}
c_{s}^{2}\equiv \frac{\left \langle\delta P_{a}\right \rangle}{\left \langle\delta \rho_{a}\right \rangle}=\frac{k^{2}/(4m_{a}^2a^{2})}{1+k^{2}/(4m_{a}^2a^{2})}.
\end{equation} 

\begin{figure}
    \centering
    \includegraphics[trim={0.6cm 0.6cm 0.2cm 0.2cm},clip,width=0.9\linewidth]{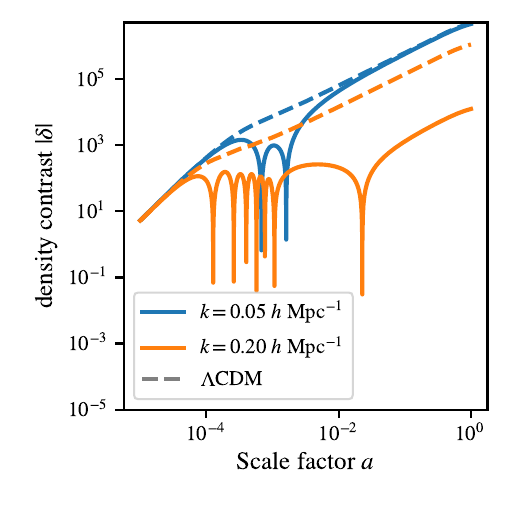}
    \caption{The growth of perturbations on small scales is suppressed in ULA models as can be seen for the $k=0.2 h~\rm{Mpc}^{-1}$ mode shown. On larger scales perturbation growth locks on to the $\Lambda$CDM solution at late times. The evolution shown here is for $m_a=10^{-26}$eV ULAs.}
    \label{fig:ula_mode_evolution}
\end{figure}

Deep in the horizon and for ULA-dominated gravitational potentials, an approximate second-order EOM can be obtained for perturbations \cite{Hu:2000ke,Chavanis:2011uv,Marsh:2015xka,Urena-Lopez:2015gur,Cookmeyer:2019rna,Hui:2021tkt}:
\begin{align}
    \ddot{\delta}_{a}+2H\dot{\delta}_{a}+\left[\frac{k^{2}c_{s}^{2}}{a^{2}}-4\pi G \rho_{a}\right]\delta_{a}=0.\label{eq:eom_2order}
\end{align} Here, dots represent derivatives with respect to coordinate time. In Eq.~(\ref{eq:eom_2order}), we can clearly see the competition between ULA pressure and self-gravity. If $k\ll k_{\rm J}$ [where the ULA Jeans scale is $k_{J}=(16\pi G a^{4}\rho_{a})^{1/4}$], we expect DM-like perturbation growth, whereas if $k\gg k_{\rm J}$, we expect oscillation rather than growth. The numerical solution to the full EOMs for individual modes (with arbitrary amplitudes) is shown in Fig.~\ref{fig:ula_mode_evolution} and bears out these expectations. 

To obtain the time-dependent matter power-spectrum $P(k,a)$ needed to compute kSZ signatures, we use the \textsc{AxionCAMB} code \cite{Hlozek:2016lzm}. For $a\leq a_{\rm osc}$, Eq.~(\ref{eq:kghom}) is solved numerically, with Eqs.~(\ref{eq:rhosfhom}) and (\ref{eq:Psfhom}) applied to determine the ULA contribution to the Friedmann equation
\begin{align}
\Delta\left( \mathcal{H}^{2}\right)=\frac{8 \pi G a^{2} \rho_{a}}{3}.
\end{align} The initial value of $\phi_{0}$ is chosen (as described in Ref. \cite{Hlozek:2014lca}) to yield the desired relic density of ULAs. 

Initially, perturbations are evolved using Eqs.~(\ref{eom:gexact_cont}) and (\ref{eom:gexact_euler}), with appropriate contributions to the metric function $h_{L}$ given by Eqs.~(\ref{eq:source_a}) and (\ref{eq:source_b}).  Once $a>a_{\rm osc}$, the scaling $\rho_{a}\propto a^{-3}$ is used (along with $P_{a}\simeq 0$), matching $\rho_{a}$ to its value at $a=a_{\rm osc}$. In this regime, the EFA equations [Eqs.~(\ref{eq:efa_cont}) and (\ref{eq:efa_euler})] are used to evolve perturbations (using $\delta P_{a}\simeq c_{s}^{2}\delta \rho_{a}$), with fluid variables continuously matched at the transition. 

Of course, the definition of $a_{\rm osc}$ is somewhat arbitrary, and a more general choice $m=nH$ could be used (the prescriptions of Refs. \cite{Urena-Lopez:2015gur} are formally equivalent to the EFA, as shown in Ref. \cite{Cookmeyer:2019rna}). Ultimately there is a trade off between improving the accuracy of the WKB approximation and decreasing the integration time available for numerical transients to dissipate. This issue is discussed extensively in Ref. \cite{Hlozek:2014lca}.

\textsc{AxionCAMB} may be used to compute the power spectra of CMB anisotropies and the matter power spectrum, defined by
\begin{align}
\left \langle \delta(\mathbf{k},a)\delta^{*}(\mathbf{k}',a)\right \rangle=\left(2\pi\right)^{3}\delta^{(3)}(\mathbf{k}-\mathbf{k}'
)P_{m}(\mathbf{k},a),
\end{align}where matter includes baryons, CDM, and ULAs in the range of $m_{a}$ values considered here.  

\section{kSZ signatures in ULA models}
\label{sec:signature}
There are in principle two approaches to observing the kSZ signature. The first is to directly search for the additional small-scale anisotropies produced by the kSZ effect using only CMB data. The second is to cross-correlate CMB maps with tracers of foreground structure.

The linear theory power spectrum of the additional small scale anisotropies induced in the CMB by the kSZ effect is given by the OV power spectrum (see Ref.  \cite{1986ApJ...306L..51O}). In Sec. \ref{sec:ov}, we derive the OV power spectrum in the presence of ULAs, computing it numerically in Sec. \ref{sec:forecast_ov}.

Pursuing the second approach, pairwise velocities of galaxy clusters can be estimated using kSZ-induced shifts to the CMB temperature along cluster sight lines. This approach was used in the first detection of the kSZ effect (see Ref.  \cite{Hand:2011ye,Hand:2012ui}). Using CMB observations from the Atacama Cosmology Telescope and galaxy clusters identified in the SDSS III Baryon Oscillation Spectroscopic Survey, the kSZ effect was detected at $2.9 \sigma$ significance, and subsequently with significance as high as $5.4\sigma$ by subsequent observational efforts, using various combinations of data from the Baryon Oscillation Spectroscopic Survey (BOSS), the South SPT, the Dark Energy Survey, the \textit{Planck} satellite, and others \cite{Hernandez-Monteagudo:2015cfa,Soergel:2016mce,DeBernardis:2016pdv,Li:2017uin,AtacamaCosmologyTelescope:2020wtv,Kusiak:2021hai,Calafut:2021wkx}. It has been shown that using pairwise velocities inferred using such kSZ observations, next-generation LSS and CMB observations will be sensitive to a number of beyond $\Lambda$CDM scenarios (see Refs. \cite{Mueller:2014nsa,Bhattacharya:2007sk} for the effect of dark energy/modified gravity or Ref. \cite{Mueller:2014dba} for probes of massive neutrinos). Indeed kSZ detections of $20\to 50\sigma$ could be be possible using DESI and Advanced ACT/S4 data \cite{Flender:2015btu}.  We derive the relevant expressions for the kSZ signature of ULA models in Sec. \ref{sec:mpv_theory}. 

\subsection{Ostriker-Vishniac effect in ULA models}
\label{sec:ov}
Our derivation of the Ostriker-Vishniac power spectrum for cosmological models with scale-dependent growth closely follows the formalism presented in Ref. \cite{Jaffe:1997ye} for a CDM cosmology but is valid in a more general context, including ULA DM, as well as for neutrinos or novel dark-energy components (whose clustering is highly suppressed). We begin with Eq.~(\ref{eq:ddtkz}) and introduce the visibility function
\begin{equation}
g(\chi)=\overline{n_e}(\chi) \sigma_T a(\chi) e^{-\tau(\chi)},
\end{equation}
in order to write
\begin{equation}\label{eq:t_anisotropies}
\frac{\Delta T}{T}= - \int \D\chi \ g(\chi) \ \bm q(\chi \hat{\bm{r}}, a) \cdot \hat {\bm r}.
\end{equation}
where $\bm q(\bm{\chi}, a) = [1+\delta(\bm{\chi},a)]\bm v(\bm{\chi}, a)$ is the momentum density expressed in terms of the density contrast $\delta(\bm{\chi},a)$. From now on we continue in Fourier space. A derivation of the Fourier transform of $\bm{q}$ is given in Appendix \ref{app:OV_derivation}. The bulk velocity depends directly on $\dot \G(k, a)$, the derivative of $\mathcal{G}(k,a)$ with respect to physical time. 

When projecting along the line of sight, any contribution of Fourier modes $\bm k$ along the line of sight must approximately cancel for small-scale modes, due to the presence of many peaks and troughs along the line of sight \cite{Jaffe:1997ye}. The contribution of the lowest-order expression $\mathbf{q}(\bm{\chi},a)\simeq \tilde{\bm{v}}(\chi,a)$ to Eq.~(\ref{eq:t_anisotropies}) thus integrates to $0$, because $\tilde{\bm v}(\bm{k}, a) \propto \bm k$. At second-order,  however, we have contributions of the form
$\int d\chi g(\chi)\int d^{3}\bm{k}' \tilde{\delta}(\bm{k}')\tilde{\bm{v}}(\bm{k}'-\mathbf{k})\cdot \hat{\bm{r}}$, as a result of the convolution theorem. Since the modes include wave vectors $\bm{k}'$ with significant components \emph{orthogonal} to the line of sight $\hat{\bm{r}}$, the second-order OV effect does not vanish. A lengthy but straightforward calculation, then yields

\begin{widetext}
 \begin{equation}
 \begin{split}
 \tilde{\bm q}_\perp(\bm k, a)= \frac{i a H(a)}{2}  \int  \frac{\D^3 \bm k'}{(2 \pi)^3}& \tilde{\delta_0}(\bm k')  \tilde{\delta_0}(\bm k- \bm k') \frac{\G(|\bm k-\bm k'|, a)}{\G_0(|\bm k-\bm k'|)}\frac{\G(k', a)}{\G_0(k')}\\ 
&\times \left[\dve{\ln \G}{\ln a}{k',a} \left(\frac{\bm k'}{k'^2} - \frac{\bm k (\bm k \cdot \bm k')}{k^2 k'^2}\right)+ 
\dve{\ln \G}{\ln a}{|\bm k-\bm k'|, a} \left(\frac{- \bm k'}{|\bm k - \bm k'|^2}+\frac{\bm k(\bm k \cdot \bm k')}{k^2 |\bm k - \bm k'|^2}\right) \right],
 \end{split}
 \end{equation}
\end{widetext}as shown in Appendix \ref{app:OV_derivation}. We have used the fact that $\mathcal{G} d\mathcal{G}/dt = \mathcal{G}^2 H(a) d\ln {\mathcal{G}}/d\ln{a}$ to obtain expressions in terms of scale factor $a$ rather than physical time.

It follows from the Limber approximation (see e.g., Ref.  \cite{loverde:2008prd}) that the power spectrum of the induced anisotropies is approximately given by
\begin{equation}\label{eq:ov_projection}
C_\ell=\int \frac{\D \chi}{\chi^2}P_\perp\left(\frac{\ell+\frac{1}{2}}{\chi}, a\right)g^2(\chi).
\end{equation}
In this expression, $P_\perp(k, a)$ is the power spectrum of the projection of $\tilde{\bm{q}}$ onto the line of sight. By expanding $\mean{\tilde{\bm q}_\perp(\bm k_1, a) \cdot \tilde{\bm q}^*_\perp(\bm k_2, a)}$, we show in Appendix \ref{app:OV_derivation} that
\begin{equation}
P_\perp(k, a) = \frac{ a^2 H^2(a)}{8 \pi^2}
S(k,a)\label{eq:vish_answer}
\end{equation}
[where $S(k,a)$ is referred to as the Vishniac power spectrum in the literature], which in contrast to $\Lambda$CDM has a time dependence

\begin{widetext}
	\begin{equation}
	\label{eq:vish_spectrum}
	\begin{split}
	S(k,a)=k \int_0^\infty \D y \int_{-1}^{1} \D x& P_0(k\sqrt{1- 2 x y +y^2})P_0(k y)\frac{\G^2(k\sqrt{1-2xy+y^2},a)}{\G^2_0(k\sqrt{1-2xy+y^2})} \frac{\G^2(k y,a)}{\G^2_0(ky)} \\
&\times \frac{1-x^2}{1-2xy+y^2}\left[\dve{\ln \G}{\ln a}{ky,a}\left(1- 2 x y +y^2\right) - \dve{\ln \G}{\ln a}{k\sqrt{1-2xy+y^2},a} y^2\right]^2
	\end{split}
	\end{equation}
\end{widetext}

This expression gives the power spectrum of secondary CMB anisotropies in the presence of ULAs and other species that induce scale-dependent growth beyond $\Lambda$CDM, and could thus be applied to determine how neutrinos and other light relics affect OV observables. 

In the limit of late-time scale-independent growth, the scale-dependent function $\mathcal{G}(k,a)\to D(a)$ (the standard $\Lambda$CDM growth function, which captures late-time structure formation) and all time-dependent terms may be factored out of the integral in Eq.~(\ref{eq:vish_spectrum}). The Vishniac power spectrum $S(k)$ then approaches the standard expression in Ref.  \cite{Jaffe:1997ye}. This can be simply understood by examining Eq.~(\ref{eq:velocity_kspace}), because if $\mathcal{G}(k,a)\to D(a)$, the scale and time dependence of $\tilde{\mathbf{v}}$ becomes significantly simpler. We assess in Sec.~\ref{sec:forecast_ov} whether these departures from the pure $\Lambda$CDM case are detectable using present and planned CMB experiments and LSS surveys.

\subsection{Mean pairwise velocity spectra in ULA models}\label{sec:mpv_theory}
For collisionless particles (e.g. DM particles or galaxies) pair conservation implies that (see Refs. \cite{1977ApJS...34..425D,Sheth:2000ff})\footnote{Strictly speaking, Eq.~(\ref{eq:pair_conservation}) is derived from the collisionless Boltzmann equation, which must be modified for wave DM. However, Eq.~(\ref{eq:pair_conservation}) holds for halos once they form, and our key results, Eqs.~(\ref{eq:mpv_averaged})-(\ref{eq:mass+volume_averaged_halo_correlation_derivative}), are still valid, as the halo model can still be used to relate halo density-correlation functions $\xi_{h}$ to $P_{0}^{\rm lin}(k,a)$ and $\mathcal{G}(k,a)$. We note, however, that Eq.~(\ref{eq:pair_conservation}) should not be interpreted as directly describing the evolution of the pairwise velocity of density fluctuations in the ULA field. Rather, the equation describes the velocity field of a limiting construct, a population of unbiased, low halo-mass tracers, as well as biased, heavy tracers of a single mass.}
\begin{equation}\label{eq:pair_conservation}
\dv{(1+\bar{\xi})}{\ln a} = -\frac{3v_{12}}{H r}\left[1+\xi\right].
\end{equation}

Here $\xi$ and $\bar{\xi}\equiv 3/(4\pi r^{3})\int_{0}^{r}4\pi r^{'2}dr'\xi(r')$ are the real-space correlation function and its volume average, respectively; $v_{12}$ is the average pairwise velocity of particles; $H$ and $a$ are the Hubble parameter and the scale factor, respectively; and $r$ is the interparticle separation. Through Eq.~(\ref{eq:pair_conservation}), $v_{12}$ can be predicted using perturbation theory and the halo model \cite{Bhattacharya:2007sk,Mueller:2014nsa,Mueller:2014dba}. ULAs would alter the growth of structure (as discussed in Sec. \ref{sec:struc_form}), thus modifying the velocity statistics predicted by Eq.~(\ref{eq:pair_conservation}).

Observationally, we are interested in the pairwise velocities of galaxy clusters, which are identified observationally in galaxy surveys. These may be estimated
by rewriting Eq.~(\ref{eq:t_anisotropies}), taking the small optical depth limit ($\tau\sim 10^{-5} \ll 1$, valid for galaxy clusters) and applying it to a single cluster sight line. The minimum variance estimator over multiple cluster sight lines in a survey is then \cite{Hand:2011ye,Planck:2013rgv,Planck:2015ywj,Hernandez-Monteagudo:2015cfa,DeBernardis:2016pdv,Li:2017uin}
\begin{equation}
    \hat{v}_{12}(r,a)=\frac{c \hat{p}_{\rm kSZ}(r,a)}{{\tau}T_{\rm CMB}},\label{eq:estmast}
\end{equation}where ${\tau}$ is the mean optical depth to a galaxy cluster and assumed not to vary significantly between clusters and $\hat{p}_{\rm kSZ}(r,a)$ is the mean pairwise momentum estimator, given by
\begin{equation}
    \hat{p}_{{\rm kSZ}}=-\frac{\sum_{i<j}\left(\delta T_{i}-\delta T_{j}\right)c_{ij}}{\sum_{i<j}c_{ij}^2}.\label{eq:pest}
\end{equation}Here $\delta T_{i}$ is the kSZ-induced CMB temperature anisotropy, while $c_{ij}$ is a geometric factor given by \cite{Ferreira:1998id}
\begin{equation}
    c_{ij}\equiv \frac{\left(r_{i}-r_{j}\right)\left(1+\cos{\theta}\right)}{2}\sqrt{r_{i}^2+r_{j}^2-2r_{i}r_{j}\cos{\theta}},
\end{equation}where $r_{i}$ and $r_{j}$ are the comoving distances to the relevant clusters and $\theta$ is their angular separation on the sky. 
If multi frequency data are available, internal linear combination techniques may be used to remove the tSZ effect from data and generate maps that contain the primary CMB and kSZ effect only, as in Refs. \cite{Planck:2015ywj,Ma:2017ybt}. Spatial filtering techniques (e.g., aperture photometry \cite{Ferraro:2014cha,Planck:2015ywj}) leveraging the known $\ell$ dependence of the primary CMB power spectrum can be used to remove the primary CMB anisotropy contribution to $\delta T_{i}$ \cite{Planck:2015ywj}. Also, individual cluster contributions are suppressed due to the averaging in Eq.~(\ref{eq:pest}) \cite{DeBernardis:2016pdv,Li:2017uin}. Once $v_{12}$ is extracted from the data, it can be compared with theoretical predictions to test hypotheses like ULA DM, among others \cite{Mueller:2014dba, Mueller:2014nsa}.

We now summarize the theoretical prediction for the cluster mass-averaged pairwise velocity $v(r)$ obtained from the predicted halo-correlation function. Each cluster represents a dark matter halo with some mass $M$. We will thus work in terms of the halo correlation function $\xi_h$. The cluster samples are typically selected for halo masses in some range $M_{\rm{min}}$ to $M_{\rm{max}}$. Averaging over halos of different masses in the sample, we can write the predicted mean pairwise velocity as

\begin{equation}\label{eq:mpv_averaged}
\begin{split}
v(r)\equiv \meanM{v_{12}}_m=&- H r\frac{\left \langle{\D\bar{\xi}_h/\D\ln a}\right \rangle_{m}}{3\left[1+ \meanM{\xi_h}_m\right]}.
\end{split}
\end{equation}
We derive this result in more detail in Appendix \ref{app:mpv_derivation}.

The mass-averaged halo correlation function is given by
\begin{equation}\label{eq:mass_averaged_halo_correlation}
\begin{split}
	\meanM{\xi_h}_m =&\frac{1}{2 \pi^2} \int k^2 \D k j_0(k r) \frac{\G^2(k,a)}{\G_0^2(k)}P_0^{\rm{lin}}(k)\mathcal{B}^2(k,a),
\end{split}
\end{equation} 
while the mass-averaged derivative of the volume-averaged correlation function is given by
\begin{equation}\label{eq:mass+volume_averaged_halo_correlation_derivative}
\begin{split}
\meanM{\dv{\bar{\xi}_h}{\ln a}}_m =&\frac{3}{\pi^2 r^3} \int_{0}^{r} \D r' {r'}^2 \int k^2 \D k j_0(k r') \\ &\times \left[ \dv{\ln \G}{\ln a} \frac{\G^2(k,a)}{\G^2_0(k)} P^{\rm{lin}}_0(k) \mathcal{B}(k,a) \mathcal{N}(k,a)\right].
\end{split}
\end{equation}

The functions $\mathcal{B}(k,a)$ and $\mathcal{N}(k,a)$ are given in terms of the halo bias $b(M, a)$, the halo mass function $n(M,a)$ and the Fourier transform of the real-space window function $\widetilde{W}(x)$ by
\begin{equation}\label{eq:mass_averaged_halo_bias}
	\mathcal{B}(k,a) = \frac{1}{\bar{n}(a)} \int_{M_{\rm{min}}}^{M_{\rm{max}}} \D M\ n(M, a) b(M, a) \widetilde{W}\left[k R(M)\right],
\end{equation}
and
\begin{equation}
\begin{split}\label{eq:mass_averaged_window}
\mathcal{N}(k,a)=&\frac{1}{\bar{n}(a)}\int_{M_{\rm{min}}}^{M_{\rm{max}}} \D M\ n(M,a) \widetilde{W}[k R(M)],
\end{split}
\end{equation}where the total halo number density is given by
\begin{equation}
    \overline{n}=\int_{M_{\rm min}}^{M_{\rm max}} n(M,a) dM
\end{equation}

It should be noted that, while Eqs.~(\ref{eq:mass+volume_averaged_halo_correlation_derivative}) and (\ref{eq:mass_averaged_window}) are similar to relevant expressions in Refs. \cite{Bhattacharya:2007sk, Mueller:2014dba, Mueller:2014nsa}, they differ in detail. We point out in Appendix \ref{app:mpv_derivation} that the expressions presented in those references exhibit unphysical behavior on small scales, biasing velocities dramatically on those scales and by as much as 16\% even on scales larger than about 20 $h^{-1}$ Mpc, possibly modifying forecasts for neutrino mass sensitivity using the kSZ effect. Our results are in agreement with Ref. \cite{Sheth:2000ff} for the case of scale-independent growth and in the absence of a window function. 

We obtain the halo mass function and bias using the semianalytic excursion set formalism [also known as the extended Press-Schechter (EPS) formalism] \cite{Sheth:1999mn}. In this model, dark matter halos are assumed to form in regions where linear growth crosses the threshold for self-similar spherical collapse. Using the statistics of Gaussian random fields and cosmological power spectra, the halo mass function is obtained. Rare peaks in the density field typically form on top of long-wavelength perturbations, and are thus more clustered (and thus \emph{biased}) than the underlying density field. The EPS model may be used to compute this bias.

The EPS halo bias is given approximately by \cite{Sheth:1999mn,Mueller:2014nsa}
\begin{equation}
b(M,a) = 1 + \frac{\delta_c^2 - \sigma_M^2(a=1)}{\sigma_M(a=1) \sigma_M(a) \delta_c},\label{eq:stbias}
\end{equation}
where $\delta_c \approx 1.686$ is the critical fractional overdensity for self-similar spherical collapse \cite{Gunn:1972sv} and $\sigma^2_M(a)$ is the variance of the matter density field smoothed on the characteristic scale associated with a cluster of mass $M$ at a scale factor $a$,
\begin{equation}
    \sigma_{M}^{2}(a)=\frac{1}{2\pi^2}\int dk k^{2} \tilde{W}^{2}(kR)P(k,a).\label{eq:sm}
\end{equation} Here $P(k,a)$ is the power spectrum at scale factor $a$.

For the halo mass function $n(M,a)$ we employ the analytic Press-Schechter approximation \cite{Press:1973iz}, which predicts that the halo mass function is given by
\begin{equation}
   n(M,a)=\sqrt{\frac{2}{\pi}}\frac{\overline{\rho}_{\rm DM}\delta_{c}}{M\sigma_M}\left|\frac{d\left(\ln{\sigma}_{M}\right)}{dM} \right|e^{-\frac{\delta_c^2}{2\sigma_M^2}},\label{eq:ps}
\end{equation}
where $n(M,a) dM$ is the number density of halos with masses in the interval $M\to M+ dM$ and $\overline{\rho}_{\rm DM}$ is the average DM mass density.\footnote{The mass function used here includes scale-dependent linear growth self-consistently, but does not include ellipsoidal collapse \cite{Sheth:1999mn}, the impact of scale-dependent growth on excursion-set barrier crossing (e.g. Refs. \cite{Du:2016zcv}), or the impact of quantum pressure on self-similar spherical collapse itself \cite{Magana:2012xe,Sreenath:2018ple}. Such issues are discussed in Refs. \cite{Marsh:2016vgj,dentler2021fuzzy} or for warm dark matter in Refs. \cite{Smith:2011ev,2012MNRAS.424..684S,Schneider:2013ria}, but are unlikely to affect our results beyond a factor of order unity, as e.g., in Ref. \cite{Du:2016zcv}.}

It has been shown that nonlinear structure in models with suppressed small-scale growth is most accurately captured by sharp $k$-space filters \cite{Schneider:2014rda}. We thus choose the window function $\widetilde{W}(x)$ such that $\widetilde{W}(x)=1$ if $x\leq 1$ and $\widetilde{W}(x)=0$ if $x>1$. We map from the halo mass $M$ to the filter length-scale $R$ using the expression $M=4\pi(\alpha R)^{3}\overline{\rho}_{\rm DM}/3$ where $\overline{\rho}_{\rm DM}$ is the mean DM density, and $\alpha\simeq 2.5$ is a factor fit to simulations \cite{Schneider:2014rda}. This factor is required because sharp-$k$ filters do not correspond uniquely to a well-defined $M$ value (due to broad support at many radii).

ULAs affect these theoretical predictions in a number of ways. They suppress the present-day linear power spectrum $P_{0}^{\rm lin}(k)$ as well as the growth function $\mathcal{G}(k,a)$ for scales $k>k_{J}$ within the ULA Jeans scale \cite{Hu:2000ke,2012PhRvD..86h3535P,Marsh:2015xka,Cookmeyer:2019rna,Hui:2021tkt}. Additionally, by suppressing small-scale structure, they increase the bias of nonlinear structures [see, e.g., Eq.~(\ref{eq:stbias})], while decreasing the number counts of smaller mass halos, as indicated by Eqs.~(\ref{eq:sm}) and (\ref{eq:ps}). 

\section{Is the ULA kSZ signature detectable?}\label{sec:detectable}
\subsection{Using the Ostriker-Vishniac power spectrum}\label{sec:forecast_ov}
In order to numerically obtain the Ostriker-Vishniac power spectrum, we output the present-day power spectrum $P_0(k)$ and the scale-dependent growth function $\G(k,a)$ using \textsc{AxionCAMB} \cite{Hlozek:2014lca}, a version of the standard cosmological Boltzmann code \textsc{CAMB} \cite{lewis:2000} that has been modified to include the impact of ULAs and output the mode evolution and $d\ln {\mathcal{G}}/d\ln{a}$. 

We then numerically evaluate the integral in Eq.~(\ref{eq:ov_projection}) to obtain predictions for the $C_{\ell}^{\rm TT}$ contributions from the kSZ effect in the presence of ULAs. We precompute and interpolate Eq.~(\ref{eq:vish_spectrum}) using 128-point Gaussian quadrature on a regular grid in $\ln(k)$ and $a$, using again Gaussian quadrature to evaluate the projection integral in Eq.~(\ref{eq:ov_projection}). Some details of the numerical methods used are discussed in Appendix \ref{app:num}.

\begin{figure}
	\includegraphics[trim={0.5cm 0.5cm 0.3cm 0.2cm},clip,width=\linewidth]{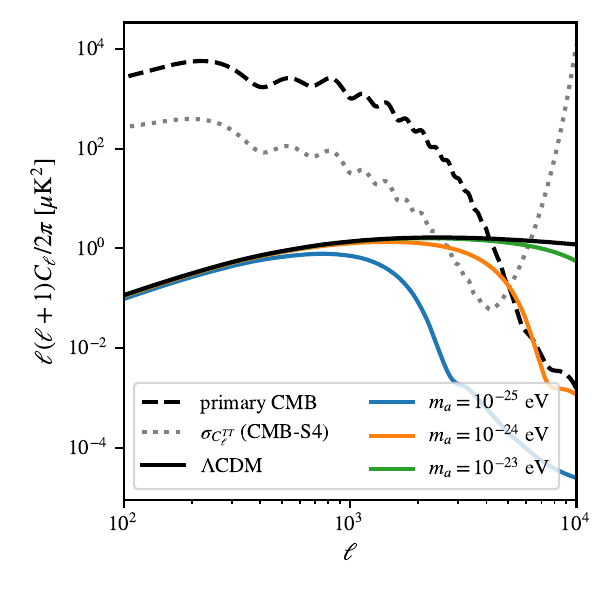}
	\caption{Ostriker-Vishniac power spectrum for cosmologies in which the total dark matter is made up of $m_a=10^{-25}$ eV, $m_a=10^{-24}$ eV, or $m_a=10^{-23}$ eV axions. For comparison the signal expected from $\Lambda$CDM model is shown. We also show the power spectrum of primary CMB fluctuations and the one-sigma uncertainty expected from a CMB-S4-like survey. The uncertainty is dominated by cosmic variance at low $\ell$ and by instrument sensitivity at large $\ell$.\label{fig:ov_power}}
\end{figure}

The results of our computations are shown in Fig.~ \ref{fig:ov_power}. We observe that the suppression of small-scale structure in the presence of axions translates into a suppression of the OV signal relative to $\Lambda$CDM. The suppression scale is set by the axion mass. Figure~\ref{fig:ov_power} also shows the primary CMB signal and the expected uncertainty for a CMB-S4-like survey. Our estimates for the S4 uncertainties are based on Refs. \cite{Knox:1995dq,He:2015msa}. We see that the typical fractional kSZ fluctuation $\Delta T\approx T_{\rm CMB}\times 10^{-2}$, justifying a perturbative treatment of the OV effect on the scales of interest.

\begin{figure*}
    \centering
    \includegraphics[width=\textwidth]{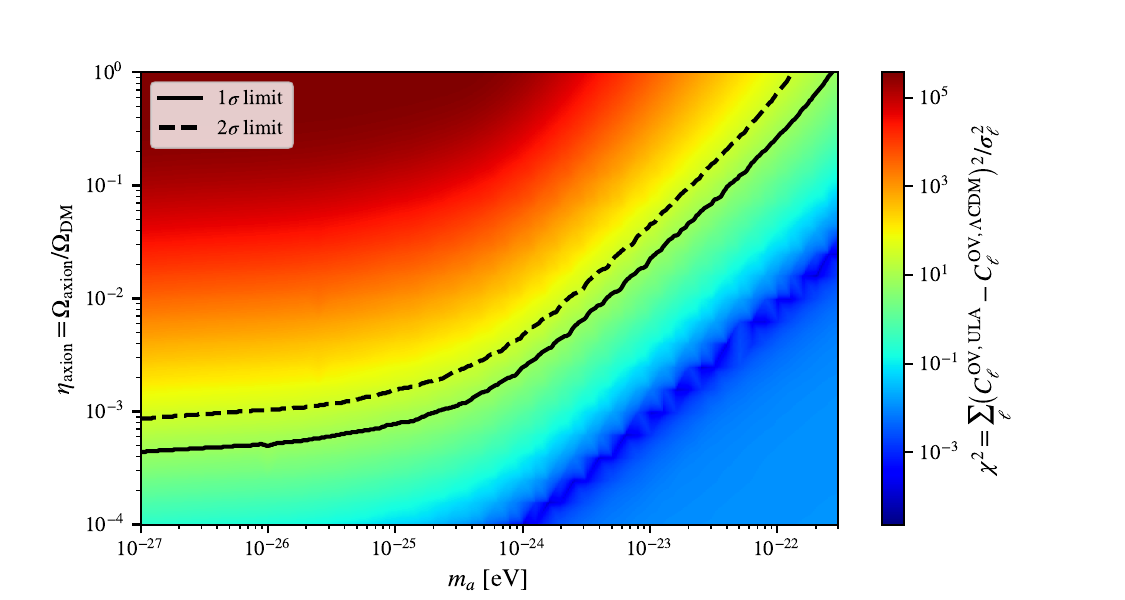}
    \caption{Rough forecast of Ostriker-Vishniac sensitivity to ULA dark matter. We show the $\chi^2$ with which any deviation from the $\Lambda$CDM prediction would be detected. The 1$\sigma$ (or 2$\sigma$) detection thresholds shown as solid (dashed) lines are estimated by requiring $\chi^2/\rm{df}=1$ ($2^2$). We assume $\rm{df}=7$ (6 $\Lambda$CDM parameters as well as the ULA abundance, $\eta_a$). The features visible in the high mass/low abundance region, well below the detection threshold, are a consequence of numerical noise.}
    \label{fig:ov_chisq}
\end{figure*}

For $\ell\lesssim3000$, the OV signal will be inaccessible due to cosmic variance and for $\ell\gtrsim5000$ even a S4-like survey will not provide the instrumental sensitivity to observe the OV signal directly. This leaves a range around $\ell\simeq4000$ in which the signal may be observed. We compute the $\chi^2$ between the OV-induced $C_{\ell}$s and standard $\Lambda$CDM predictions, showing the result as a heat map in Fig.~\ref{fig:ov_chisq}. We see that values $\eta_{\rm axion}\simeq 10^{-3}$ are detectable in the range $10^{-27}~{\rm eV} \lesssim m_{a}\lesssim 10^{-25}~{\rm eV}$. We see that the data are sensitive to $\eta_{a}\simeq 1$ up to $m_{a}\simeq 10^{-22}~{\rm eV}$, and so it is possible that the OV effect is sensitive to ULAs in the true FDM window, where they  could compose all of the DM. Of course this requires extremely accurate subtraction of the primary CMB, using TT measurements at low-$\ell$ or E-mode polarization anisotropies over a broad range of $\ell$.

Additionally, we note that the curves in Fig.~\ref{fig:ov_power} were obtained using our second-order perturbative results Eqs.~(\ref{eq:ov_projection}) and (\ref{eq:vish_spectrum}) and were computed in the approximation of instantaneous reionization. Additionally, the detailed shape of the ULA-induced modifications to the OV signature will have degeneracies with $\Lambda$CDM parameters. Our sensitivity estimate from ULA-induced changes to the OV effect is thus likely to be overly optimistic.

A more realistic treatment would include the impact of the topology of reionization (the ``patchy reionization" signature, quantified by a bubble power spectrum for ionized regions), as described in Ref. \cite{Calabrese:2014gwa}.  Such a computation would also include the impact of ULAs in delaying reionization (see Ref. \cite{Bozek:2014uqa} for a discussion) and their effect on the bubble power spectrum (see Ref. \cite{Roncarelli:2017cwe} for an example of how neutrinos alter the nature of patchy reionization and the resulting OV/kSZ observables). Our results for the magnitude and future sensitivity of OV signatures in ULA models should be taken as a provisional indication that they might be experimentally detectable, motivating more elaborate modeling in future work. 

\subsection{Using mean pairwise velocity Spectra}\label{sec:vel}

\begin{figure*}
	\centering
	\includegraphics[width=\linewidth]{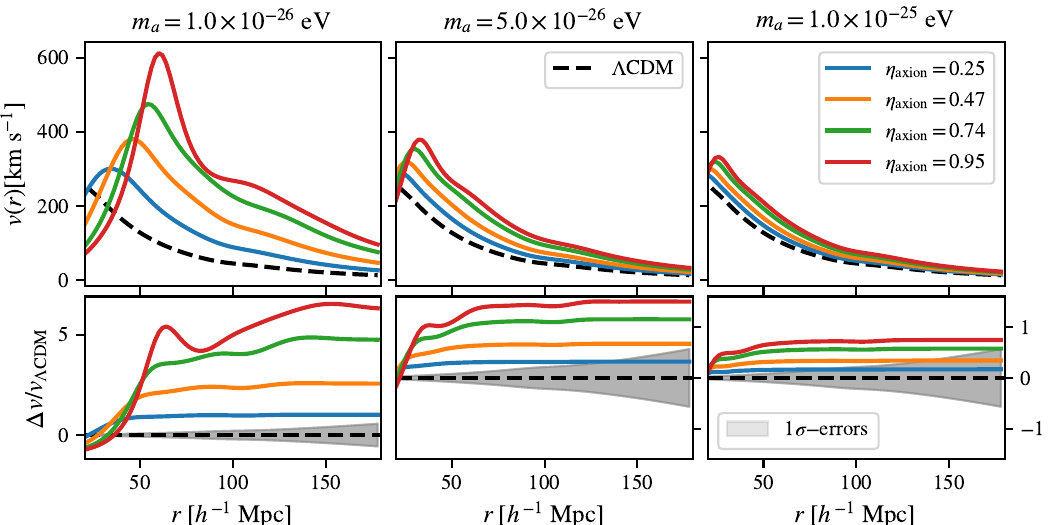}
	\caption{Predicted mean pairwise velocities at $z=0.15$ for three different axion masses and different abundances. Predictions for a $\Lambda$CDM model are shown for comparison. The gray bands show the velocity uncertainty, computed using the diagonal elements of the covariance expected for an S4-like survey. We adopt the covariance expression presented in Ref. \cite{Mueller:2014dba} with minor adjustments to include our modifications to the mean pairwise velocity [see Eqs.~(\ref{eq:cov_measurement}) and (\ref{eq:cov_cosmic})].}
	\label{fig:vab}
\end{figure*}

We now turn to the mean pairwise velocity approach. As in the previous section we obtain present-day density fluctuation variables and their time evolution using \textsc{AxionCAMB}. We then compute the expected mean pairwise velocity spectra according to the expressions presented in Sec. \ref{sec:mpv_theory}. We employ Convolutional Fast Integral Transforms as implemented in \textsc{mcfit}\footnote{\url{https://github.com/eelregit/mcfit/}} to evaluate the relevant integrals presented in Sec. \ref{sec:mpv_theory} and Gaussian quadrature for the bias integrals involving finite limits [Eq. (\ref{eq:mass_averaged_halo_bias}) and (\ref{eq:mass_averaged_window})]. The results are shown in Fig.~ \ref{fig:vab}. A simpler summary is depicted in Fig.~ \ref{fig:dmp}.

\begin{figure}
    \centering
    \includegraphics[trim={0.1cm 0.1cm 0.5cm 0.1cm},clip,width=\linewidth]{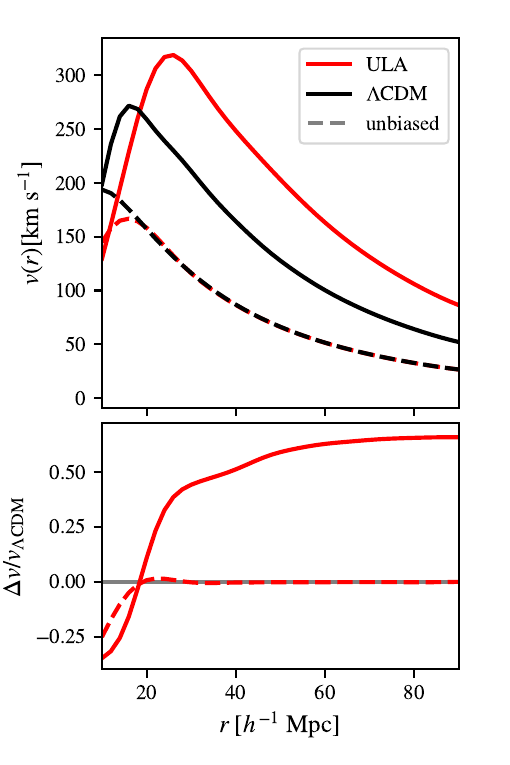}
    \caption{\emph{Top panel:} Mean pairwise velocities of the matter density field (dashed) and galaxy field (solid) in comparison. We compare a model with $m_a=5 \times 10^{-26}$ eV and $\eta_a=0.5$ (red) to a $\Lambda$CDM model (black). The velocities of the matter density field (dashed lines) are suppressed by axions on small scales and approach $\Lambda$CDM on larger scales, while galaxy pairwise velocities exhibit enhancement at large separations due to large bias. \emph{Bottom panel:} Fractional differences between ULA and $\Lambda$CDM pairwise velocity signatures, for an unbiased tracer of the DM density field (dashed) and halos (solid), respectively.}
    \label{fig:dmo}
\end{figure}

For small comoving separations $r$, mean pairwise velocities are suppressed in the presence of axions relative to a $\Lambda$CDM model. The suppression scale increases with decreasing axion mass and increasing axion abundance. At large separations, axions lead to an enhancement of observed pairwise velocities. This is due to the fact that the same massive clusters are higher-$\sigma$ peaks of the cosmological density field than in $\Lambda$CDM models. They are thus rarer and exhibit stronger clustering (larger bias), causing an enhancement at large $r$. This effect is visualized in Fig.~\ref{fig:dmo}. We observe that if galaxy bias is neglected (i.e., computing the mean pairwise velocity of the matter density field), velocities in the presence of axions are suppressed on small scales and approach the $\Lambda$CDM prediction on large scales. Using the same cosmological model but now including halo bias (i.e. computing the galaxy pairwise velocities) then leads to the enhancement on large separations (as also noted in Refs. \cite{Bauer:2020zsj,Lague:2021frh}). 

\subsection{Forecast for ULA abundance sensitivity of pairwise velocity spectra}

As the impact of ULA dark matter on mean pairwise velocities is comparable to the error bars (e.g.,  Fig.~\ref{fig:vab}) of forthcoming experiments (determined from their experimental covariance matrix, modeled as discussed in Sec.~\ref{sec:covar}), it is plausible that ULA DM is detectable using the kSZ effect. 

We thus proceed with a standard Fisher-matrix forecast (following the formalism developed in Refs. \cite{Tegmark:1996bz,Bond:1997wr,Eisenstein:1998hr}), in which the likelihood of a model (specified by a set of parameters) is obtained in the limit of small deviations from the fiducial model, yielding an approximately Gaussian model-parameter posterior. 

Given some axion mass $m_a$, we consider $\Lambda$CDM as a model specified by a parameter-space vector $\mathbf{\Theta}$ consisting of five of the six $\Lambda$CDM parameters as well as the axion abundance $\eta_a=\Omega_{\rm{axion}/}\Omega_{\rm DM}$,
$$\bm{\Theta}=\left(\Omega_{\rm{DM}} h^2, \Omega_b h^2, h, n_s, A_s, \eta_a\right).$$ 
For the fiducial cosmology $\bm{\Theta}_\text{fid}$, we assume that all $\Lambda$CDM parameters take the fiducial values obtained by the Planck Collaboration in their final full-mission analysis (2018) \cite{Aghanim:2018eyx}, that is, \emph{total} (including ULAs) dark-matter density, $\Omega_{\rm DM} h^2=0.120$, baryon density $\Omega_{\rm b}h^2=0.0224$ scalar spectral index $n_s=0.965$, $h=0.674$, and $\ln{(10^{10}A_{s})}=3.04$.

Mean pairwise velocities are insensitive to the optical depth to reionization, and we hence choose to fix it to its best-fit value from Ref. \cite{Aghanim:2018eyx}, $\tau_{\rm reion}=0.054$.  In addition to these six cosmological parameters, we follow Refs.~ \cite{DES:2018gxz,DES:2021zxv} and consider a set of nuisance parameters $b_i$ that scale the bias $\mathcal{B}\to b_{i} \mathcal{B}$ independently in the $i$th redshift bin. This accounts for the uncertainty (due to a variety of baryonic effects) in the mapping from observed galaxy masses to dark-matter halo masses, as a function of $z$. Unless otherwise noted, we marginalize over these these parameters to obtain all the results below.

The kSZ Fisher matrix is then given by a sum over redshifts and comoving radii, \begin{equation}\label{eq:fisher_matrix_expression}
F_{i j} =\sum_k^{N_z}\sum_{m,n}^{N_r} \pdv{v(r_m,z_k)}{\theta_i} \mathbf{C}^{-1}_{r_m, r_n, z_k, z_k}\pdv{v(r_n,z_k)}{\theta_j}.
\end{equation}
Here $N_z$ and $N_r$ are the number of redshift and radial bins respectively.  Here, $\mathbf{C}^{-1}_{r_m, r_n, z_k, z_k}$ is the appropriate element of the inverse-covariance matrix given by Eqs. (\ref{eq:cov_measurement}) and (\ref{eq:cov_cosmic}). Forecast uncertainties on individual parameters (labeled by the index $i$) after marginalization over the others are then given by $\sigma_{i}=\sqrt{\mathbf{F}^{-1}_{ii}}$, where $\mathbf{F}^{-1}$ denotes the inverse of the Fisher matrix.

In order to determine the minimum axion fraction which could be detected given some axion mass $m_a$, we consider a range of fiducial axion abundances $\eta_a$ between $10^{-4}$ and $0.95$. Twenty values are chosen to span this range logarithmically, with $20$ more values chosen to make sure $1$ and $2-\sigma$ detection thresholds are well resolved in sensitivity plots.\footnote{We find that the derivatives obtained via finite difference rule are contaminated by numerical noise for step sizes smaller than about 5\%. The use of one sided difference rules also introduces spurious signatures for all sufficiently large step sizes. Consequently, we are unable to properly probe $\eta_a=1$.}  For the axion mass $m_a$, $41$ values are chosen, distributed logarithmically to cover the domain from $10^{-27}~{\rm eV}\to 10^{-23}~{\rm eV}.$ As noted in Refs. \cite{Amendola:2005ad,Hlozek:2014lca,Hlozek:2016lzm,Hlozek:2017zzf}, the posterior probability of $m_{a}$ is highly non-Gaussian, and so Fisher analysis is of limited use for $m_{a}$ itself. It is thus easiest to follow Refs. \cite{Amendola:2005ad,Hlozek:2014lca,Hlozek:2016lzm,Hlozek:2017zzf} and consider $m_{a}$ as a fixed parameter. At each value of $m_{a}$, we conduct a Fisher sensitivity forecast with respect to $\eta_{a}$. The detection threshold is  obtained as the minimum axion abundance for which the forecast 1$\sigma$ (or 2$\sigma$) uncertainties on $\eta_a$ are smaller than $\eta_a$ itself. 

Similarly to Ref. \cite{Mueller:2014dba}, we consider three different CMB survey stages. SII represents currently available data, SIII-like surveys will become available in the near future, and SIV represents long-term prospects. The survey specifications and expected uncertainties on the measured pairwise velocities are summarized in Tables \ref{tab:kSZ_surveys} and \ref{tab:uncertainties} respectively. We consider a DESI-type galaxy survey \cite{Aghamousa:2016zmz}. A spectroscopic galaxy sample can of course be arbitrarily divided into $z$ bins without changing the fundamental information content of the sample. For consistency with Ref.  \cite{Mueller:2014nsa,Mueller:2014dba}, however, we choose $N_{z}=5$ $z$-bins. We note that we could have considered a different number of bins, making $z$-evolution of the velocity field more manifest, but 
with smaller numbers of pairs in each bin such that total signal-to-noise (and ULA sensitivity) is unchanged. 

\begin{table}
	\begin{center}
		\caption{Reference survey specifications used to model SII, SIII, and SIV (reproduced from Ref.  \cite{Mueller:2014nsa}).\label{tab:kSZ_surveys}}
		\begin{tabular}{l l c c c}
			\hline\hline
			\multicolumn{2}{c}{} & \multicolumn{3}{c}{ Survey Stage }
			\\	\cline{3-5}
			Survey & Parameters \  & \ SII\footnote{Currently available CMB/LSS surveys such as ACTPol and SDSS BOSS.} \ & \ SIII\footnote{Near-term survey generations (e.g. AdvACTPol) and SDSS BOSS dataset.} \ & \  SIV\footnote{Long-term survey prospects such as CMB-S4 combined with a LSS dataset such as DESI.} \    \\ \hline
			CMB&    $\Delta T_{\mathrm{instr}}$ ($\mu K \mathrm{arc\ min}$) \ & 20 &  7  & 1
			\\
			Galaxy &   $z_{\mathrm{min}}$ &0.1&0.1&0.1
			\\
			&  $z_{\mathrm{max}}$  \ &0.4&0.4&0.6
			\\
			& No. of $z$ bins, $N_z$   \   &3&3&5
			\\
			&  $M_{\mathrm{min}}$ ($10^{14}M_\odot $) \ & $1 $ & $1$ &  $0.6$
			\\
			\multicolumn{2}{l}{Overlap area (1000 deg$^2$)} \ & 4 & 6& 10
			\\ \hline \hline
		\end{tabular}
	\end{center}
\end{table}

\begin{table}
	\begin{center}
		\caption{Uncertainties for different survey stages. The table is reproduced from Ref. \cite{Mueller:2014nsa}.}
		\begin{tabular}{lccccccc}
			\hline\hline\\
			\multirow{2}{*}{Parameter}&\multirow{2}{*}{\makecell{Survey\\Stage\footnote{Survey parameters for different stages are provided in Table \ref{tab:kSZ_surveys}.}}}& \multicolumn{5}{c}{ Redshift bin} \\ \cline{3-7}
			& & $0.15$ & $0.25$ & $0.35$ & $0.45$  &  $0.55$ \\ \hline
			$(\Delta \tau/\tau)^2$ &&\multicolumn{5}{c}{ 0.15}\\
			$\sigma_{\tau}$ (km/s)&&\multicolumn{5}{c}{ 120}\\ \hline
			$\sigma_{\mathrm{instr}}$ (km/s)& SII& 290 &  440 &   540&-&-\\
			& SIII &100&  150 & 190 &-&-\\
			& SIV& 15&  22& 27&  34&  42\\ \hline
			$\sigma_{v}$(km/s) & SII&310&  460 & 560 &-&-\\
			& SIII&160& 200&   230 &-&-\\
			& SIV&120&  120& 120& 120&  130\\\hline
		\end{tabular}
		\label{tab:uncertainties}
	\end{center}
\end{table}
\label{sec:covar}
We adopt the covariance prescription presented in the Appendix of Ref. \cite{Mueller:2014nsa}, modifying the expressions there with our expressions for $v(r)$ and neglecting the subdominant, non-Gaussian contribution. The covariance matrix for the mean pairwise velocity spectra has three dominant components: one from the measurement uncertainty, one due to cosmic variance, and one due to sampling noise. 
We assume that the measurement uncertainty is uncorrelated between different radial separation ($r$) and redshift ($z$) bins and only contributes to the diagonal elements of the covariance matrix \cite{Mueller:2014nsa,Mueller:2014dba,Bhattacharya:2007sk}
\begin{equation}\label{eq:cov_measurement}
C_\text{measurement}\vert_{r_n, r_m, z_j, z_k}= \frac{2\sigma_v^2}{N_\text{pair}} \delta_{mn}\delta_{j k}.
\end{equation}

Here, $\sigma_v=\sqrt{\sigma_{\rm instr}^{2}+\sigma _\tau^{2}}$ is the uncertainty on the velocity measurement, including both the direct measurement error $\sigma_{\rm instr}$ and the variance in $v$, $\sigma _\tau^{2}$, induced by the variance in the optical depth, through the scaling $\tau\propto v^{-1}$, shown in Eq.~(\ref{eq:estmast}). We thus have $\sigma_\tau= v \Delta \tau/\tau$.  Both contributions are estimated in Table \ref{tab:uncertainties}.

The number of cluster pairs, denoted $N_\text{pair}$, is given by  \begin{equation}
N_\text{pair} = \frac{\bar{n}(z) V_s(z)}{2} \left(4 \pi \int_{r}^{r+\Delta r} \bar{n}(z)\left[1+\xi_h(r,z)\right] r^2 dr\right).
\end{equation}
The average number density of clusters at a given redshift $z$ is
\begin{equation}
  \bar{n}(z)  = \int_{M_{\rm{min}}}^{M_{\rm{max}}} \D m\ n(m, z),
\end{equation} where $V_s(z)$ is the survey volume as a function of scale factor. The halo sample is taken to have lower and upper mass limits $M_{\rm{min}}$ and $M_{\rm{max}}$. We can see that $\bar{n}(z) V_s(z)$ is the total number of clusters in the survey at a given $z$. The number of clusters in a spherical shell of inner radius $r$ and outer radius $r+\Delta r$ (where $\Delta r$ is the radial bin width) around a given cluster is $4 \pi \int_{r}^{r+\Delta r} \bar{n}(z)\left[1+\xi_h(r,z)\right] r^2 dr$. Thus, the product of these two factors gives the number of pairs, and in order to avoid double counting, we divide by $2$ which gives the expression above. Assuming that $\xi_h$ is approximately constant over the interval from $r$ to $r+\Delta r$, we have
\begin{equation}
N_\text{pair} = \frac{\bar{n}^2(z) V_s(z) V_\Delta(r)}{2} \left[1+\xi_h(r,z)\right],
\end{equation}
where $V_\Delta$ is the volume of the radial bin.

The contribution from cosmic variance and shot noise is given by\begin{widetext}
\begin{equation}\label{eq:cov_cosmic}
\begin{split}
\Big[C_\text{cosmic} + C_\text{shot}\Big]_{r_m, r_n, z_j, z_k}=&\frac{4\delta_{j k}z^2_k}{\pi^2 V_s(z_j)} \left(\frac{H^2(z_j) }{\left[1+ \xi_h(r_m)\right]\left[1+ \xi_h(r_n)\right]}\right) \left(\dve{\ln D}{\ln \left[1+z\right]}{z_j}\right)^2 \\
&\times \int dk \Bigg[ \left(P(k,z_j)\mathcal{B}(k, z_j) \mathcal{N}(k,z_j)  + \frac{1}{n(z_j)}\right)^2 \times W_\Delta(k, r_m)W_\Delta(k,  r_n)\Bigg],
\end{split}
\end{equation}\end{widetext}
where $W_\Delta$ is
\begin{equation}
W_\Delta(k, r) = 2\left\{\frac{r^3\mathcal{W}(kr)- (r+\Delta r)^3\mathcal{W}[k(r+\Delta r)]}{(r+\Delta r)^3 -r^3}\right\},
\end{equation}
and 
$$\mathcal{W}(x) = \frac{2 \cos x + x \sin x}{x^3}.$$ The factors of $W_{\Delta}$ and $\mathcal{W}$ arise from Fourier transforms and integrals over real-space covariance expressions for pairs of clusters with radial separations within a fixed bin with width $\Delta r$ (and the resulting Bessel functions). The usual $\Lambda$CDM growth function $D(a)$ is defined by the relation $P^{\rm{lin}}(k,a) = P^{\rm{lin}}_0(k) D^2(a)/D^2(a=1)$, and captures late-time scale-independent growth, as is the case for the fiducial model. 

\begin{figure}
	\centering
	\includegraphics[trim=0cm 1.5cm 0cm 1.5cm, clip]{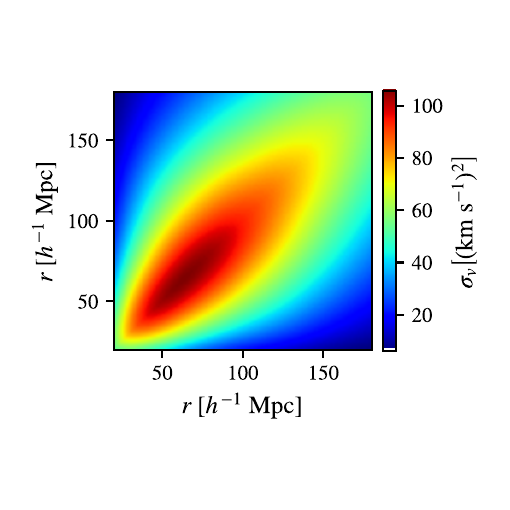}
	\caption{Full covariance for mean pairwise velocity spectra constructed from a SIV-like survey at redshift $z=0.15$ [see Eqs. (\ref{eq:cov_measurement}) and (\ref{eq:cov_cosmic})]. The individual components contribution to the covariance are shown in Fig.~\ref{fig:covaraince_contrib}.}
	\label{fig:covaraince}
\end{figure}

\begin{figure}
	\centering
	\includegraphics[width=\columnwidth]{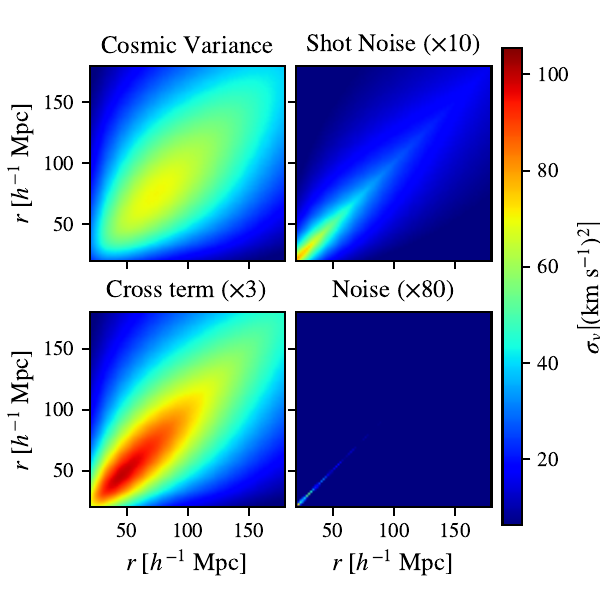}
	\caption{Contributions to the total mean pairwise velocity covariance in the lowest redshift bin centered on $z=0.15$ [see Eqs. (\ref{eq:cov_measurement}) and (\ref{eq:cov_cosmic})]. Top left: Cosmic Varaiance; top right: shot noise ($\times 10$); bottom left: shot noise/cosmic variance cross term ($\times 3$); and bottom right: measurement uncertainty mostly due to scatter in cluster optical depth ($\times 80$).}
	\label{fig:covaraince_contrib}
\end{figure}

The resulting covariance matrix in the lowest redshift bin centered on $z=0.15$ for a SIV survey is shown in Fig.~ \ref{fig:covaraince}. Additionally, the different contributions to the covariance are detailed in Fig.~\ref{fig:covaraince_contrib}. We see there that cosmic variance dominates along the diagonal at all scales, with secondary contributions from shot noise. The contribution due to scatter in the cluster optical depth is negligible compared to other contributions.

The approximate error bars shown in Fig.~\ref{fig:vab} are obtained by fixing $z$ and then taking $\sqrt{\mathbf{C}_{r_m,r_m,z_k,z_k}}$. At large $r$, the covariance flattens due to the fact that the measurement error drops off with the increasing number of pairs in a volume, while the window function $W_{\Delta}$ asymptotes to a constant. The signal $v(r)$ itself falls off at very large separations. As a result, there is a rise in the fractional error at large $r$.

\begin{figure*}
    \centering
    \includegraphics{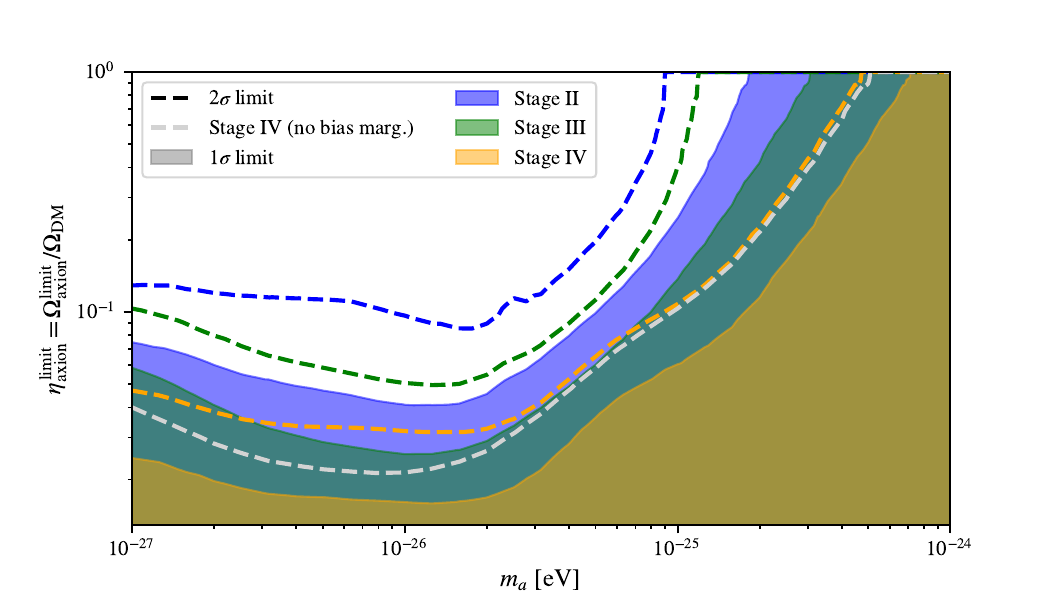}
    \caption{Forecasted detection sensitivity in $\eta_a=\Omega_a/\Omega_\mathrm{DM}$ as a function of the ULA mass $m_{a}$ for SII, SIII and SIV surveys as defined by Ref.  \cite{Mueller:2014nsa}. Regions above the dotted lines (or shaded areas) would be detectable at $2\sigma$ (or $1\sigma$). The maximum mass that can be probed at the 2$\sigma$ level with SII and SIII surveys is of the order $m_a\simeq 10^{-25}\rm{eV}$ and up to $m_a \simeq 5 \times 10^{-25}$ eV with SIV. When we do not marginalize over the bias nuisance parameters $b_i$ the constraints are tightened in the mass region below about $m_a \simeq 3\times10^{-26}$eV. }
    \label{fig:fisher}
\end{figure*}

\begin{figure*}
    \centering
    \includegraphics[width=\textwidth]{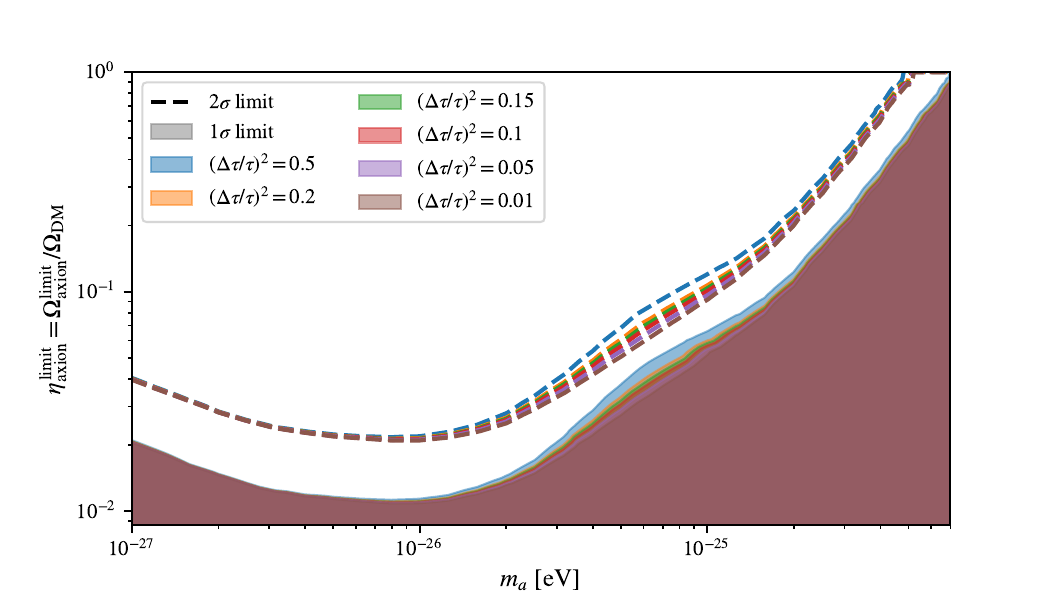}
    \caption{Forecasted detection sensitivity in $\eta_a=\Omega_a/\Omega_\mathrm{DM}$ as a function of $m_{a} $ for an SIV survey as defined by Ref. \cite{Mueller:2014nsa}, for a number of different priors on the mean cluster optical depth $\tau$. Regions above the dotted lines (or shaded areas) would be detectable at $2\sigma$ (or $1\sigma$). }
    \label{fig:fisher_taudep}
\end{figure*}

\begin{figure*}
    \centering
    \includegraphics[width=\textwidth]{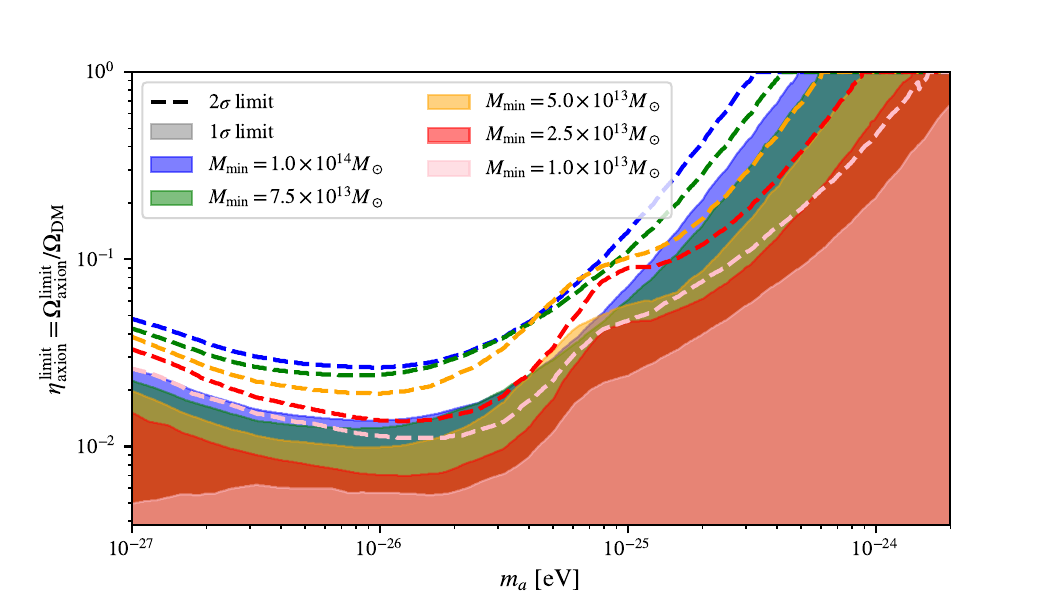}
    \caption{Forecasted detection sensitivity in $\eta_a=\Omega_a/\Omega_\mathrm{DM}$ as a function of $m_{a}$ for an SIV survey as defined by Ref. \cite{Mueller:2014nsa}, for different minimum cluster masses. As above regions above the dotted lines (or shaded areas) would be detectable at $2\sigma$ (or $1\sigma$). Here, we do not marginalize over uncertainties in the bias. Doing so degrades the constraints obtained with lower minimum masses more strongly, partially eliminating any gains made by including lower mass clusters. The main improvement is the ability to probe higher axion masses.}
    \label{fig:fisher_mMindep}
\end{figure*}

We obtain numerical derivatives with respect to our six cosmological parameters by finite differencing using a five-point rule and adopting the step sizes suggested by Ref.  \cite{Eisenstein:1998hr} for the five $\Lambda$CDM parameters. We test different step sizes between $1\%$ and $40\%$ in $\eta_a$ and find excellent convergence across the entire axion mass range within the few percent level for all step sizes $\gtrsim 5\%$.

The  minimum axion abundance that may be detected at 1$\sigma$ (2$\sigma$) significance via mean pairwise velocities alone is shown in Fig.~\ref{fig:fisher}, obtained by evaluating Eq.~(\ref{eq:fisher_matrix_expression}). We can see that for axion masses well below $m_a \simeq 10^{-25}$ eV the axion abundance could be strongly constrained by kSZ observations alone (to the $\sim 10\%$ level with SII or III and at the percent level with SIV). The sensitivity worsens rapidly with increasing $m_{a}$. The maximum mass that can be probed with a SII and SIII survey is around $3\times10^{-26}$ and $6\times 10^{-26}$ eV, respectively. With SIV, this increases to about $2\times10^{-25}$ eV. We also show that there is a slight dependence of forecasted detection limits on our knowledge of the expected halo bias. Neglecting the bias nuisance parameters $b_i$ tightens the constraints for axion masses $m_a\lesssim 3\times 10^{-26}$~eV.

In Figs.~\ref{fig:tri1}-\ref{fig:tri4}, we show the degeneracies between $\eta_{a}$, the standard cosmological parameters, and the bias parameters $b_{1}$, $b_{2}$, $b_{3}$, $b_{4}$, and $b_5$, for several fiducial parameter sets of $m_{a}$ and $\eta_{a}$. These figures are generated using a methodology described in Appendix \ref{sec:degen}. We note that there are strong degeneracies within the bias model. There are also strong degeneracies within the pairs $\{n_{s},b_{j}\}$ and the pairs $\{A_{s},b_{j}\}$. This level of degeneracy is responsible for the difference between the constraints obtained when marginalizing over vs neglecting bias nuisance parameters.

We additionally also tested the impact of varying assumptions on the scatter in the cluster optical depth, which arises due to the variance in the cluster population, not measurement error. In our fiducial analysis, we adopt $(\Delta \tau/\tau)^2=0.15$, similarly to Ref. \cite{Mueller:2014dba}, leading to an optical-depth induced uncertainty in the mean pairwise velocity of $\sigma_\tau=120\rm{km/s}$ (see Table \ref{tab:uncertainties}). We tested $(\Delta \tau/\tau)^2$ values between $0.001$ and $0.8$ without major impact on detection limits, as shown in Fig.~\ref{fig:fisher_taudep}.

\begin{figure}
	\includegraphics{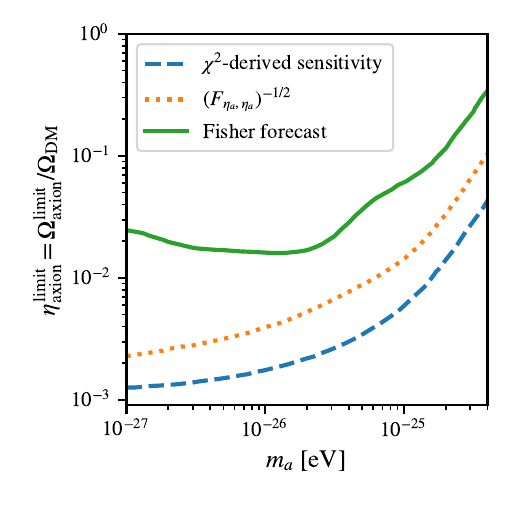}
	\caption{Comparison of $\chi^2$-derived sensitivity level with Fisher-matrix result. We would expect the $\chi^2$-derived sensitivity to agree approximately with the inverse square root of the diagonal element of the Fisher matrix corresponding to the axion abundance. We find this to be the case up to a approximately mass-independent factor of $\sim$2. Our Fisher forecast is a conservative estimate of the detection limits. This is likely due to the linear expansion of $v(r)$ around $\Lambda$CDM values.}
	\label{fig:chi_sq}
\end{figure}

We also explored the promise of future survey efforts with much lower minimum halo masses. We recomputed Fisher matrices with a number of $M_{\rm min}$ values. We found that the sensitivity of pairwise velocity estimators alone could
improve by a factor of $\sim 3$ in $\eta_{a}$ if $M_{\rm min}\simeq 10^{13}M_{\odot}$, as shown in Fig.~\ref{fig:fisher_mMindep}.

To verify our results we conduct a $\chi^2$-analysis of the $\eta_{a}$ sensitivity of the kSZ effect. In this approach, the likelihood for the observables is treated as Gaussian, but the full (nonlinear) dependence of observables on model parameters is used. In other words, we went beyond the Fisher approximation to critically assess its validity.

We fixed all parameters except the axion abundance to their fiducial values. For a single varying parameter ($\eta_{a}$), this approach is in principle exact, and the predicted $1\sigma$ uncertainty should agree approximately with the inverse square root of the $\eta_a$ diagonal element of the Fisher matrix. The results are shown in Fig.~\ref{fig:chi_sq}, and indeed if only $\eta_a$ is varied, the $\chi^2$ and Fisher-level sensitivities agree, up to a nearly mass-independent factor of $\sim$2. This difference results from the assumption of Gaussian posteriors and the linear expansion of $v(r)$ around fiducial $\Lambda$CDM values. The overall trend is that our forecasts are likely more conservative than a complete future data analysis.

\subsection{Combining results from mean pairwise velocity spectra with primary CMB observations}

We combine and compare our results with primary CMB observations and CMB lensing measurements as they are expected from a CMB-S4-like survey. In addition to the six cosmological parameters we vary in our kSZ analysis we also include the optical depth to the CMB in the forecast for the primary CMB observations and CMB lensing.  
We compute the CMB Fisher matrix using the \textsc{OxFISH} code \cite{Allison:2015qca}, by varying the axion parameters in combination with the other five primary parameters. 

As described above and in  Ref. \cite{Hlozek:2016lzm}, for fixed axion mass $m_a$, we assume a range of fiducial axion fractions, given that the current constraints from cosmology are only upper limits. The step size assumed in a Fisher matrix forecast is a key factor in determining the balance between the accuracy of the derivatives and numerical noise. To account for this, we vary the step size assumed in a range from $\delta_\Theta/\Theta_* = 0.2, 0.1, 0.05, 0.01$ for a given fiducial value $\Theta_*,$ to check for the stability of the final Fisher error $\sigma_\Theta.$

We make the following assumptions about the analysis of future CMB-S4 data combined with \emph{Planck}. For the lowest multipoles $2<\ell<30$ we use a modified \emph{Planck} configuration that mimics a prior of $\sigma(\tau)=0.01$ on the optical depth. For the range $30<\ell<2500$ we model the \textit{Planck} HFI instrument but only on 20\% of the sky to remove ``double counting" of CMB-S4 numbers on the same sky area. Finally, we include the CMB-S4 noise modeled as a Gaussian component with a beam of $1$ arc minute and a noise level of $1\rm{\mu K\ arc\ min},$ included via the Knox formula \cite{Knox:1995dq}, \begin{equation} 
N_{\alpha\alpha}= (\Delta_\alpha)^2 \exp\left(\frac{\ell(\ell+ 1)^2\theta_\mathrm{FWHM}^2}{8 \ln 2}\right).\end{equation}

\begin{figure}
	\includegraphics[width=\columnwidth]{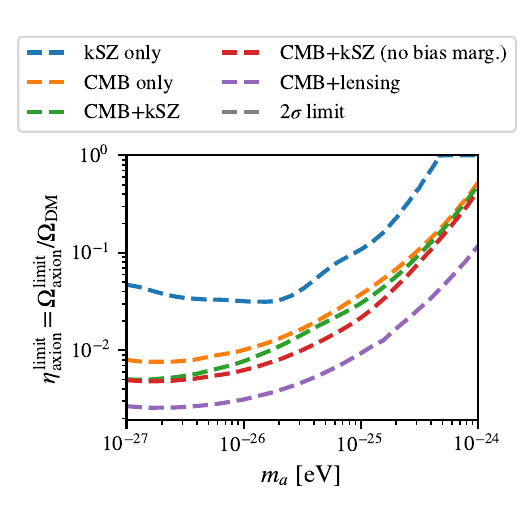}
	\caption{Primary CMB and kSZ observations will be sensitive to  axion fractions down to about $\sim$0.5\% at $2\sigma$ significance for masses below $10^{-26}$eV. We show the $2\sigma$ detection limits for a combination of DESI and CMB-S4.}
	\label{fig:combined}
\end{figure}

The polarization noise is a factor of $\sqrt{2}$ larger than the temperature noise. Both are included between $30< \ell < 4000.$ In addition, we include the lensing deflection power spectrum from $30<\ell< 3000$. We compare the runs with and without adding information from the lensing deflection reconstruction in Fig.~\ref{fig:combined}. The lensing deflection, which couples the modes in temperature and polarization to reconstruct the lensing potential, is computed using the Hu and Okamoto quadratic-estimator formalism \cite{Hu:2001kj}.

We find that combining kSZ and CMB observations allows sensitivity to an abundance of $\sim$0.5\% below $m_a=10^{-26}$eV. This is an improvement over observations of the primary CMB alone as shown in Fig.~\ref{fig:combined}. When marginalization over bias nuisance parameters is taken into account, the improvement over CMB-only constraints diminishes with increasing axion mass. 

This sensitivity level is competitive with the combination of primary CMB and CMB lensing to within a factor of order unity, roughly consistent with the comparative sensitivity of the same observables to the neutrino mass, as discussed in Ref. ~\cite{Mueller:2014dba}. Further improvements are likely possible using large, photometric samples, higher $n$-point functions of the reconstructed velocity field, lower $M_{\rm min}$ values, or foreground tracers, like field galaxies or neutral gas line-intensity maps \cite{Sato-Polito:2020cil}. Additionally, the combination of kSZ observations with the primary CMB can provide a valuable cross-check on CMB and CMB lensing results.

\resub{It is interesting to consider these forecasts in the context of the sensitivity of LSS observables at the level of $2$-pt correlations, perhaps as measured using a photometric galaxy survey such as that planned for the Large Synoptic Survey Telescope (LSST) \cite{2009arXiv0912.0201L,Bechtol:2019acd}. Preliminary forecasts by some of us and others \cite{lstrott} indicate that in the mass window $10^{-27}~{\rm eV}\lesssim m_{a}\lesssim 10^{-25}~{\rm eV}$, LSST's galaxy survey alone should be sensitive to $5\times 10^{-2}\lesssim  \eta_{\rm axion}\lesssim 10^{-1}$ comparable to pairwise measurements of the kSZ alone. LSST would manifest largely mass-independent sensitivity to $\eta_{\rm axion}$ as high as $m_{a}\sim 10^{-23}~{\rm eV}$, so the primary strength of kSZ data is to offer comparable sensitivity for a sub-dominant but non-negligible component of the dark sector.}

\resub{If limits to neutrino abundances are a reliable guide, the inclusion of priors to the LSST projections from CMB acoustic-scale anisotropy measurements could improve sensitivity to $\eta_{\rm axion}$ by a factor of $\sim 0.2$ reduction in error bar \cite{Abdalla:2007ut}. In parallel, the same priors would \emph{also} improve kSZ sensitivity by another order of magnitude, though both of these statements are crude estimates that await a proper future forecast. Galaxy power spectrum and kSZ observables are thus on their own comparably sensitive to ULAs.}

\resub{Galaxy power spectra and pairwise velocity signatures have different dependencies on unknown bias factors, $b$, specifically scaling as $\sim b^{2}$ and $\sim b$ respectively, and it is thus likely that these distinct data sets will prove complementary by breaking each others' degeneracies. Weak lensing is likely to be comparably sensitive to this new physics, but manifests distinct systematics (e.g. galaxy alignment, image point-spread function measurement errors) \cite{Mandelbaum:2017jpr}, making combined probes necessary to robustly detect new physics.}

\resub{At the moment, there are constraints to ULA DM from the absorption spectra of high-$z$ quasars, known as the Lyman-$\alpha$ forest \cite{Kobayashi:2017jcf,Armengaud:2017nkf,Irsic:2017ixq,Nori:2018pka,Rogers:2020ltq}, imposing a limit of $\eta_{\rm axion} \lesssim 0.2$ for $m_{a}\lesssim 10^{-21}~{\rm eV}$. Future Lyman-$\alpha$ measurements could reach an order of magnitude lower sensitivity to the absorption optical depth \cite{Aghamousa:2016zmz}, and while a ULA-specific forecast does not yet exist, it could be that this offers an additional factor of $\sim 10$ improvement in sensitivity $\eta_{\rm axion} \lesssim 0.2$ for $m_{a}\lesssim 10^{-21}~{\rm eV}$, competitive with the pairwise kSZ sensitivity level forecast in our work.}

\resub{Thinking further ahead into the future, intensity mapping efforts with the cosmological $21$-cm and other lines could offer novel probes of the linear density field.
Efforts like HIRAX \cite{Newburgh:2016mwi} and the Square Kilometer Array (SKA) \cite{Staveley-Smith:2015iva} could offer a full additional order-of-magnitude improvement in sensitivity $\eta_{\rm axion}$ for masses as high as $m_{a}\sim 10^{-24}~{\rm eV}$ \cite{Bauer:2020zsj}, but must progress to a robust $21$-cm fluctuation detection before being useful as a fundamental physics probe.}

 \section{Conclusions}
\label{sec:conc}
The next decade of cosmological observations will yield nearly cosmic-variance limited measurements of CMB polarization, as well as deep spectroscopic surveys of $\sim 10^{7}$ galaxies that facilitate ever more precise maps of cosmological large-scale structure. These measurements will improve our understanding of reionization, cluster thermodynamics, radio point sources, galaxy formation, and fundamental physics \cite{Abazajian:2019eic}. Increasingly, cosmological data will be used not only to probe the dark-sector energy budget but also its particle content. 

Ultralight axions could exist over many decades in mass and are a well-motivated candidate to compose some or all of the dark matter. Going beyond WMAP and \textit{Planck} measurements, much of the sensitivity of upcoming CMB experiments to dark-sector particle physics will be driven by secondary anisotropies, such as gravitational lensing and the kinetic Sunyaev-Zel'dovich effect \cite{Abazajian:2019eic}. 

In this work, we have computed the ULA signature on Ostriker-Vishniac CMB anisotropies imprinted after reionization, and on the pairwise cluster velocity dispersion (measured using the CMB and cluster surveys), including scale-dependent growth in a self-consistent manner. In future work, we will explore the impact of our analytic results on predictions for kSZ signatures of neutrinos. The OV signature of ULAs was found to be detectable if $\eta_{a}\gtrsim 10^{-3}$ at S4 sensitivity levels with fairly simple assumptions. Future work will examine the robustness of this signature to degeneracies with a number of reionization-related parameters and realistic subtraction of the primary CMB, as well as other relatively featureless foregrounds. Proposed futuristic small-scale efforts like CMB-HD could offer even more promising  opportunities to detect this signature \cite{Nguyen:2017zqu,Sehgal:2019ewc}. This signature seems competitive with all the LSS probes considered above, but in future efforts, we must carefully consider foregrounds and marginalization over our ignorance of the true model of reionization (which could itself be inhomogeneous) \cite{Alvarez:2020gvl}.

Using ULA linear perturbation theory and the halo model of structure formation \cite{Bardeen:1985tr,Sheth:1999mn,Sheth:1999su,Sheth:2000ii,Sheth:2000fe,Sheth:2000ff,Sheth:2001dp,Cooray:2002dia}, we found that if $10^{-27}~{\rm eV}\leq m_{a}\leq 2\times 10^{-25}~{\rm eV}$ CMB-S4 and DESI could together reveal ULA mass fractions in the range $0.002\leq \Omega_{a}/\Omega_{d}\leq 0.02$, offering comparable sensitivity to CMB lensing \cite{Hlozek:2016lzm}. In future work, it will be valuable to jointly assess lensing and kSZ observables for ULA sensitivity, in order to fully account for degeneracy breaking from these multiple observables. 

Our forecast assumed a spectroscopic redshift survey (e.g., DESI). Future photometric LSS experiments like LSST, however, will produce surveys with $10^3\to10^{4}$ times as many galaxies, while sacrificing accuracy in redshift \cite{2009arXiv0912.0201L,Bechtol:2019acd}. Although such surveys will suffer from lower signal-to-noise than comparably voluminous redshift surveys (due to washout of modes with large projections along the line of sight) \cite{Smith:2018bpn}, they have already been used for kSZ pairwise velocity detections \cite{Soergel:2016mce}; in the future, we will assess the kSZ-driven sensitivity of LSST and other photometric surveys (combined with CMB data) to ULA signatures, as well as the complementary nature of more direct measurements of the matter two-point function.

Going forward, we could build upon the halo-model techniques employed here, for example, using more accurate halo mass functions and the accompanying Sheth-Tormen bias functions \cite{Sheth:1999mn}, extending our model to properly include the effect of scale-dependent barrier crossing (as in Ref.  \cite{Du:2016zcv}). We somewhat arbitrarily included ULAs in the definition of DM used to calculate fractional density contrasts. In future efforts, we can follow the lead of Ref.  \cite{Chiang:2017vuk} for massive neutrinos, and account for the fact that some fraction of the ULA mass density will be bound and some will be unbound. Given the tremendous recent progress in numerical simulations of ULA structure formation using hydrodynamic, Schr\"{o}dinger-Poisson, and modified $N$-body solvers \cite{Schive:2014dra,Mocz:2017wlg,Nori:2018hud,Schwabe:2020eac}, it would be interesting to directly apply simulation outputs (including baryon physics where possible) in order to more realistically model kSZ observables in the presence of ULAs.

As shown in Ref. \cite{Smith:2018bpn}, a variety of statistical methods for analyzing kSZ data are equivalent to the pairwise velocity dispersion used here, as they are all fundamentally tied to the $\delta \delta v $ bispectrum \cite{Smith:2018bpn}. One such method uses the peculiar velocity field-estimator $\hat{v}$, obtained using off-diagonal correlations of the CMB temperature field and galaxy density \cite{Deutsch:2017ybc}. An advantage of this language is that it  furnishes another useful kSZ statistic, the correlation function $\langle \hat{v}(\vec{r}+\vec{x})\hat{v}(\vec{r})\rangle$ evaluated at comoving separation $\hat{x}$, leveraging four-point correlations (the trispectrum) to provide additional statistical power, potentially breaking degeneracies of cosmological parameters with bias parameters and the mean kSZ optical depth \cite{Smith:2018bpn}.

Past work on using the kSZ effect as a probe of novel physics explored its sensitivity to neutrino mass and novel (non-GR) anisotropic stress in the gravitational sector. Here, we have gone further and demonstrated the utility of the kSZ effect as a probe of the nature of dark matter. There are a variety of other theoretical possibilities related to dark matter that would also suppress structure formation, with changes in power spectra similar to ULAs, such as nonstandard baryon-DM scattering \cite{Li:2018zdm,Xu:2018efh}, neutrino-DM scattering \cite{Binder:2016pnr}, or sterile neutrino DM (see Ref.  \cite{Abazajian:2017tcc} and references therein).  Future efforts should thus establish the full sensitivity of the kSZ effect to a broad range of theoretical  dark-sector models.

\begin{acknowledgments}
D.~G. acknowledges support in part by NASA ATP Grant No. 17-ATP17-0162. G.~S.~F. acknowledges support through the Isaac Newton Studentship and the Helen Stone Scholarship at the University of Cambridge. D.~G. and G.~S.~F. acknowledge support from the Provost's office at Haverford College. G.~S.~F. acknowledges support from the KINSC Summer Scholars Fund at Haverford College. G.~S.~F. thanks Imperial College for hospitality during the conduct of some of the research described here. A.~H.~J. acknowledges support from STFC in the United Kingdom. R.~H. is a CIFAR Azrieli Global Scholar, Gravity \& the Extreme Universe Program, 2019, and a 2020 Alfred P. Sloan Research Fellowship. RH is supported by Natural Sciences and Engineering Research Council of Canada and the Connaught Fund. The work of D.~J.~E.~M was supported by the Alexander von Humboldt Foundation and the German Federal Ministry of Education and Research. D.~J.~E.~M is supported by an Ernest Rutherford Fellowship from UK STFC.

The authors acknowledge useful conversations with T.~L.~Smith, A.~v.~Engelen, R.~Sheth, A.~Kosowsky, F.~Cyr-Racine, M.~Dentler, A.~Lague, B.~Sherwin, and T.~Baldauf. The authors thank Bruce Partridge for useful conversations and a careful reading of the manuscript. DG and GSF are grateful to J.~Cammisa for assistance with the \textbf{fock} computer cluster at Haverford College. 

The land on which the Haverford College stands is part of the ancient homeland and unceded traditional territory of the Lenape people. We pay respect to Lenape peoples, past, present, and future and their continuing presence in the homeland and throughout the Lenape diaspora. 

The Dunlap Institute is funded through an endowment established by the David Dunlap family and the University of Toronto. We acknowledge that the land on which the University of Toronto is built is the traditional territory of the Haudenosaunee, and most recently, the territory of the Mississaugas of the New Credit First Nation. We are grateful to have the opportunity to work in the community, on this territory. 
\end{acknowledgments}

\onecolumngrid
\appendix

\section{Detailed derivation of Ostriker-Vishniac power spectrum}\label{app:OV_derivation}
\renewcommand\thefigure{\thesection.\arabic{figure}}    
\setcounter{figure}{0}    

For this paper, we adopt the  following Fourier conventions:
\begin{eqnarray}
\tilde{f}(\bm k)=\int \D^3\bm x e^{-i \bm k \cdot \bm x} f(\bm x),\\
f(\bm x)=\int \frac{\D^3 \bm k}{(2 \pi)^3} e^{i \bm k \cdot x} \tilde{f}(\bm k).
\end{eqnarray}  
We will start with our expression for the projected temperature anisotropies [Eq.~(\ref{eq:t_anisotropies})],
\begin{equation}
\delta T=\frac{\Delta T}{T}= - \int \D\chi \ g(\chi) \ \bm q(\chi \hat{\bm{r}}, a) \cdot \hat {\bm r}.
\end{equation}
where we have defined the momentum density $\bm q(\bm{\chi}, a) = [1+\delta(\bm{\chi},a)]\bm v(\bm{\chi}, a)$. Here, the visibility function $g(\chi)$ is the projection kernel for the field $Q(\bm \chi, a) = \bm{q}(\chi \hat{\bm{r}}, a) \cdot \hat {\bm r}$. The Fourier transform of $\bm q(\bm \chi,a)$ is given by
\begin{equation}
\tilde{\bm q}(\bm k, a)= \int \D^3\bm \chi e^{-i \bm k \cdot \bm \chi} \bm q(\bm \chi, a)=\tilde{\bm v}(\bm k, a) + \int \frac{\D^3 \bm k'}{(2 \pi)^3}  \tilde{\delta}(\bm k', a) \tilde{\bm v}(\bm k- \bm k', a),
\end{equation}
which we obtained by substituting for $\delta(\bm \chi, a)$ in terms of its Fourier transform. We could have just as easily substituted in for $\bm v(\bm \chi, a)$ and obtained
\begin{equation}
\tilde{\bm q}(\bm k, a)= \tilde{\bm v}(\bm k, a) + \int \frac{\D^3 \bm k'}{(2 \pi)^3}  \tilde{\bm v}(\bm k', a) \tilde{\delta}(\bm k- \bm k', a).
\end{equation}
For symmetry reasons, we will thus use
\begin{equation}
\tilde{\bm q}(\bm k, a)= \tilde{\bm v}(\bm k, a) + \frac{1}{2}\int \frac{\D^3 \bm k'}{(2 \pi)^3} \left[ \tilde{\delta}(\bm k- \bm k', a)\tilde{\bm v}(\bm k', a) + \tilde{\delta}(\bm k', a) \tilde{\bm v}(\bm k- \bm k', a) \right].
\end{equation}
Using Eq. (\ref{eq:velocity_kspace}), we can write this expression solely in terms of the density contrast and the growth factor
\begin{equation}
\begin{split}
\tilde{\bm q}(\bm k, a)
= \frac{i a H(a)}{k^2} \frac{\G(k, a)}{\G_0(k)} \dv{\ln \G}{\ln a} \bm k \tilde{\delta_0}(\bm k) +  \frac{i a H(a)}{2}  \int  \frac{\D^3 \bm k'}{(2 \pi)^3} & \tilde{\delta_0}(\bm k')  \tilde{\delta_0}(\bm k- \bm k') \frac{\G(|\bm k-\bm k'|, a)}{\G_0(|\bm k-\bm k'|)}\frac{\G(k', a)}{\G_0(k')} \\ 
\times& \left[\dve{\ln \G}{\ln a}{k',a} \frac{\bm k'}{k'^2}+ \dve{\ln \G}{\ln a}{|\bm k-\bm k'|, a} \frac{\bm k- \bm k'}{|\bm k- \bm k'|^2} \right].
\end{split}
\end{equation} 
As argued in the main body of this work and more rigorously shown by Ref. \cite{Jaffe:1997ye}, only modes perpendicular to the line of sight contribute appreciably to the line of sight integral, and thus the projection of $\bm q(\bm k,a)$ onto the line of sight is approximately given by $\bm q_\perp(\bm k,a)$, the projection onto the direction perpendicular to $\bm k$. We can obtain this projection by
\begin{equation}
\tilde{\bm q}_\perp(\bm k, a)=\left(\bm{I}-\frac{\bm{K}}{k^2}\right) \cdot \tilde{\bm q}(\bm k, a),
\end{equation}
where $\bm{I}$ is the identity matrix and $\bm{K}$ is a matrix, such that $K_{ij}=k_i k_j$. This yields
\begin{equation}
\begin{split}
\tilde{\bm q}_\perp(\bm k, a)= \frac{i a H(a)}{2}  \int  \frac{\D^3 \bm k'}{(2 \pi)^3}& \tilde{\delta_0}(\bm k')  \tilde{\delta_0}(\bm k- \bm k') \frac{\G(|\bm k-\bm k'|, a)}{\G_0(|\bm k-\bm k'|)}\frac{\G(k', a)}{\G_0(k')}\\ 
\times& \left[\dve{\ln \G}{\ln a}{k',a} \left(\frac{\bm k'}{k'^2} - \frac{\bm k (\bm k \cdot \bm k')}{k^2 k'^2}\right)+ 
\dve{\ln \G}{\ln a}{|\bm k-\bm k'|, a} \left(\frac{- \bm k'}{|\bm k - \bm k'|^2}+\frac{\bm k(\bm k \cdot \bm k')}{k^2 |\bm k - \bm k'|^2}\right) \right].
\end{split}
\end{equation}
The power spectrum $P_\perp(k)$ is defined by
\begin{equation}
\mean{\tilde{\bm q}_\perp(\bm k_1, a) \cdot \tilde{\bm q}^*_\perp(\bm k_2, a)}=(2 \pi)^3\delta_D(\bm k_1 - \bm k_2)P_\perp(k_1, a).
\end{equation}
From Wick's theorem, it follows that 
\begin{equation}
\begin{split}
&\mean{\tilde{\delta}_0(\bm k_1 - \bm k_1')\tilde{\delta}_0(\bm k_1')\tilde{\delta}^*_0(\bm k_2 - \bm k_2') \tilde{\delta}^*_0(\bm k_2')}\\
&=(2 \pi)^6 P_0(|\bm k_1 - \bm k_1'|)P_0(k_1')[\delta_D(\bm k_1 - \bm k_2)\delta_D(\bm k_1' - \bm k_2') + \delta_D(\bm k_1 - \bm k_2)\delta_D(\bm k_1-\bm k_1' - \bm k_2')],
\end{split}
\end{equation}
where $P_0^{\rm lin}(k)$ is the linear mass power spectrum at the present time. Therefore, we obtain
\begin{equation}
\begin{split}
\mean{\tilde{\bm q}_\perp(\bm k_1, a) \cdot \tilde{\bm q}^*_\perp(\bm k_2, a)}
=&\delta_D(\bm k_1 - \bm k_2) \frac{a^2H^2(a)}{2} \int \D^3 \bm  k_1' P_0(|\bm k_1 - \bm k_1'|)P_0(k_1') \frac{\G^2(|\bm k_1-\bm k_1'|, a)}{\G^2_0(|\bm k_1-\bm k_1'|)}  \frac{\G^2(k_1', t)}{\G^2_0(k_1')}\\
&\times \left[\dve{\ln \G}{\ln a}{k_1',a} \left(\frac{\bm k_1'}{k_1'^2} - \frac{\bm k_1 (\bm k_1 \cdot \bm k_1')}{k_1^2 k_1'^2}\right)+ \dve{\ln \G}{\ln a}{|\bm k_1-\bm k_1'|,a} \left(\frac{- \bm k_1'}{|\bm k_1 - \bm k_1'|^2}+\frac{\bm k_1(\bm k_1 \cdot \bm k_1')}{k_1^2 |\bm k_1 - \bm k_1'|^2}\right) \right]^2.
\end{split}
\end{equation}
In order to integrate over all space we change to spherical coordinates defined such that $\bm k =(k, \theta=0, \phi=0)$. Furthermore, we substitute $\theta = \cos^{-1}x $ and $k'=k y$. With these substitutions, we have $\bm k \cdot \bm k'=k^2 xy$ and $|\bm k - \bm k'|=k\sqrt{1-2xy+y^2}$. We finally find
\begin{equation}
\begin{split}
\mean{\tilde{\bm q}_\perp(\bm k, a) \cdot \tilde{\bm q}^*_\perp(\bm k_2, a)}=&(2 \pi)^3\delta_D(\bm k - \bm k_2) \frac{ a^2 H^2(a)}{8 \pi^2} k \int_0^\infty \D y \int_{-1}^{1} \D x P_0(k\sqrt{1- 2 x y +y^2})P_0(k y)\frac{1-x^2}{1-2xy+y^2}\\ 
&\times \frac{\G^2(k\sqrt{1-2xy+y^2},a)}{\G^2_0(k\sqrt{1-2xy+y^2})} \frac{\G^2(k y,a)}{\G^2_0(ky)}
\left[\dve{\ln \G}{\ln a}{ky,a}\left(1- 2 x y +y^2\right) - \dve{\ln \G}{\ln a}{k\sqrt{1-2xy+y^2},a} y^2\right]^2.\label{eq:ovfinal}
\end{split}
\end{equation}
Our expression is in agreement with Ref. \cite{Jaffe:1997ye} when the scale dependence of $\G$ is dropped.\footnote{There is, however, a difference of a factor of $2$ between the two derivations. The same difference was found in Ref. \cite{Jaffe:1997ye}, when comparing to other published results. Our expression is in agreement with the other published results.} We thus write the analog of the Vishniac $S(k,a)$ power spectrum as in Eq.~ (\ref{eq:vish_spectrum}), including additional time dependence as expressed there.
\section{Mean pairwise velocity spectra}\label{app:mpv_derivation}
\renewcommand\thefigure{\thesection.\arabic{figure}}    
\setcounter{figure}{0}    

As discussed in the body of this paper, we start with the pair conservation equation as given by Ref. \cite{1977ApJS...34..425D} and cited by Ref. \cite{Sheth:2000ff}:
\begin{equation}\label{eq:raw_mpv}
	\dv{(1+\bar{\xi})}{\ln a} = -\frac{3v_{12}}{H r}\left[1+\xi\right].
\end{equation}
Here, $\xi$ and $\bar{\xi}$ are correlation function and its volume average, respectively; $v_{12}$ is the average pairwise velocity of any two particles in the field; $H$ and $a$ are the Hubble parameter and the scale factor, respectively; and $r$ is the separation between particles. 

Consider now halos of mass $m$, a biased tracer of the matter field $\delta(\bm{x})$ smoothed with a spherically symmetric window function on some characteristic scale $R$ that depends on $m$. If the bias $b$ is linear and a function of $m$ and the scale factor only, we have
$$\delta_h^{(m)}(\bm{x}) = b(m,a) \int \delta(\bm{x})W_R(\abs{\bm{x}-\bm{y}}) \D^3\bm{x}.$$
Now, by the Fourier convolution theorem, we can write the transform of this as 
$$\tilde{\delta}^{(m)}_h(\bm{k}) = b(m,a) \tilde{\delta}(\bm{k}) \widetilde{W}_R(\abs{\bm{k}}),$$
where the tilde shall denote Fourier transformed quantities. We can define a function $\widetilde{W}(x)$ such that $\widetilde{W}(k R) = \widetilde{W}_R(k)$. 

Consider now halos with masses $m_1$ and $m_2$. The cross power spectrum at equal time is given by
\begin{equation}
\begin{split}
\mean{\tilde{\delta}_h^{(m_1)}(\bm{k})\tilde{\delta}_h^{(m_2)*}(\bm{k}')} =& b(m_1, a) b(m_2, a) \mean{\tilde{\delta}_h(\bm{k})\tilde{\delta}_h^{*}(\bm{k}')}\widetilde{W}(k R_1) \widetilde{W}(k' R_2)\\
=& (2\pi)^3 \delta_D^{(3)}(\bm{k}-\bm{k'}) b(m_1, a) b(m_2, a) P^{\rm{lin}}(k, a) \widetilde{W}(k R_1) \widetilde{W}(k' R_2).
\end{split}
\end{equation}
In the above, we have written the characteristic size of halos of mass $m_1$ and $m_2$ as $R_1$ and $R_2$, respectively.

We assume that the linear power spectrum can be written in terms of the present day power spectrum $P^{\rm{lin}}_0$ and a growth function $\G(k,a)$, which in our case depends on scale. $P^{\rm{lin}}(k,a) = P^{\rm{lin}}_0(k) \G^2(k,a)/\G_0^2(k)$. The correlation function of halos of masses $m_1$ and $m_2$ is therefore given by
\begin{equation}\label{eq:correlation}
	\xi_h^{(m_1, m_2)} = \frac{1}{2 \pi^2} \int k^2 \D k j_0(kr) \frac{\G^2(k,a)}{\G_0^2(k)} P^{\rm{lin}}_0(k)b(m_1, a)b(m_2, a)\widetilde{W}(k R_1)\widetilde{W}(k R_2).
\end{equation}
The halo bias $b(m,a)$ is given to good approximation by (see Ref. \cite{Sheth:2000ff})
\begin{equation}
b(m,a) = 1 + \frac{\delta_c^2 - \sigma_m^2(a=1)}{\sigma_m(a=1) \sigma_m(a) \delta_c},
\end{equation}
where $\sigma_m^2$ is the variance of the matter density field smoothed on some scale $R(m)$ and $\delta_c \approx 1.686$ is the critical collapse overdensity for self-similar spherical collapse \cite{Gunn:1972sv}.

If the growth of structure is scale independent, the derivative of the bias with respect to $\ln a$ is given by (see e.g., Ref.  \cite{Sheth:2000ff})
\begin{equation}
	\dv{b}{\ln a} = \dv{\ln D}{\ln a}\left[1-b(m,a)\right].
\end{equation}
We argue below that, even in the case of scale-dependent growth this derivative is well approximated on all scales of interest by 
\begin{equation}\label{eq:bias_deriv}
	\dv{b}{\ln a} = \dv{\ln \G}{\ln a}\left[1-b(m,a)\right].
\end{equation}

Taking the derivative $\D \xi_h^{(m_1,m_2)}/ \D \ln a$ yields
\begin{equation}
\begin{split}
\dv{\xi_h^{(m_1,m_2)}}{\ln a} = \frac{1}{2 \pi^2} \int k^2 \D k\ & j_0(kr) P^{\rm{lin}}_0(k)\widetilde{W}(k R_1)\widetilde{W}(k R_2)\\
\times& \Bigg[\dv{\ln a}\left(\frac{\G^2(k,a)}{\G^2_0(k)}\right) b(m_1, a)b(m_2, a)\\
&+ \frac{\G^2(k,a)}{\G^2_0(k)} \dve{b}{\ln a}{m_1, a}b(m_2, a)\\
&+ \frac{\G^2(k,a)}{\G^2_0(k)} b(m_1, a)\dve{b}{\ln a}{m_2, a}\Bigg]
\end{split}
\end{equation}
which simplifies with the help of Eq.~(\ref{eq:bias_deriv}) to
\begin{equation}\label{eq:correlation_derivative}
\begin{split}
\dv{\xi_h^{(m_1,m_2)}}{\ln a} = \frac{1}{2 \pi^2} \int k^2 \D k\ & j_0(kr) \dv{\ln \G}{\ln a}\frac{\G^2(k,a)}{\G^2_0(k)} P^{\rm{lin}}_0(k)\widetilde{W}(k R_1)\widetilde{W}(k R_2)\\
\times& \Bigg[2 b(m_1, a)b(m_2, a)\\
&+ \left[1-b(m_1,a)\right] b(m_2, a)\\
&+ b(m_1, a) \left[1-b(m_2,a)\right]\Bigg]\\
= \frac{1}{2 \pi^2} \int k^2 \D k\ & j_0(kr) \dv{\ln \G}{\ln a}\frac{\G^2(k,a)}{\G^2_0(k)} P^{\rm{lin}}_0(k) \left[b(m_1, a) + b(m_2,a) \right]\widetilde{W}(k R_1)\widetilde{W}(k R_2).
\end{split}
\end{equation}
Lastly, we still need to take the volume average of Eq. (\ref{eq:correlation_derivative}) as follows:
\begin{equation}\label{eq:correlation_derivative_volume_avg}
	\dv{\bar{\xi}_h^{(m_1,m_2)}}{\ln a} = \frac{3}{r^3} \int_0^r (r')^2 \D r' \dv{\xi_h^{(m_1,m_2)}}{\ln a}
\end{equation}

Following Eq. (\ref{eq:raw_mpv}), the average pairwise velocity of pairs of halos of masses $m_1$ and $m_2$ is then
\begin{equation}\label{eq:mpv_m1_m2}
	v_{12}^{(m_2,m_2)} =- \frac{H r}{3\left[1+ \xi_h^{(m_1,m_2)}\right]} \dv{\bar{\xi}_h^{(m_1,m_2)}}{\ln a}
\end{equation}
with $\xi_h^{(m_1,m_2)}$ and $\D \bar{\xi}_h^{(m_1,m_2)}/ \D \ln a$ given by Eqs. (\ref{eq:correlation}) and (\ref{eq:correlation_derivative_volume_avg}), respectively.

To obtain the pairwise velocity averaged over pairs of different masses in the halo sample used, we weight this by the product of the number density per unit mass of clusters of mass $m_1$ and the number density per unit mass of clusters of mass $m_2$ a distance $r$ from the former, relative to the total number density of cluster pairs in our sample separated by a distance $r$,
\begin{equation}\label{eq:weighting}
	w(r, a, m_1, m_2) = \frac{n(m_1,a) n(m_2, a) \left[1+\xi_h^{(m_1,m_2)}\right]}{\bar{n}^2(a)\left[1+ \mean{\xi_h}_m\right]}.
\end{equation}
Here, $\bar{n}(a) = \int_{M_{\rm{min}}}^{M_{\rm{max}}} \D m\ n(m, a)$ is the total number density of clusters with lower and upper mass limits $M_{\rm{min}}$ and $M_{\rm{max}}$ for the halo sample  and $\mean{\xi_h}_m$ indicating the sample-averaged halo correlation function defined by 
\begin{equation}
\begin{split}
	\mean{\xi_h}_m =& \frac{1}{\bar{n}^2(a)}\int_{M_{\rm{min}}}^{M_{\rm{max}}} \D m_1 \int_{M_{\rm{min}}}^{M_{\rm{max}}} \D m_2 \ n(m_1,a)n(m_2,a) \xi_h^{(m_1, m_2)}\\
	=&\frac{1}{2 \pi^2} \int k^2 \D k j_0(k r) \frac{\G^2(k,a)}{\G^2_0(k)} P_0^{\rm{lin}}(k)\mathcal{B}^2(k,a).
\end{split}
\end{equation}

Here, $\mathcal{B}(k,a)$ is defined as
\begin{equation}\label{eq:mass_averaged_halo_bias_app}
	\mathcal{B}(k,a) = \frac{1}{\bar{n}(a)} \int_{M_{\rm{min}}}^{M_{\rm{max}}} \D m\ n(m, a) b(m, a) \widetilde{W}\left[k R(m)\right].
\end{equation}
There is no window function in the denominator in the definition of $\overline{n}(a)$ here, in contrast with the expressions in Refs. \cite{Bhattacharya:2007sk,Mueller:2014nsa}. 

Combining now the weighting from Eq. (\ref{eq:weighting}) with Eq. (\ref{eq:mpv_m1_m2}) and integrating over $m_1$ and $m_2$, we have the  mean pairwise velocity
\begin{equation}
	V_h=\mean{v_{12}}_m =- \frac{H r}{3\left[1+ \mean{\xi_h}_m\right]} \frac{1}{\bar{n}^2}\int \D m_1 \int \D m_2 n(m_1,a) n(m_2,a)\dv{\bar{\xi}_h^{(m_1,m_2)}}{\ln a}
\end{equation}

We notice that the integral appearing here gives the sample average over the $\ln a$ derivative of the volume averaged halo correlation function
\begin{equation}\label{eq:mass+volume_averaged_halo_correlation_derivative_app}
\begin{split}
\meanM{\dv{\bar{\xi}_h}{\ln a}}_m =& \frac{1}{\bar{n}^2}\int \D m_1 \int \D m_2\ n(m_1,a) n(m_2,a)\dv{\bar{\xi}^{(m_1,m_2)}_h}{\ln a}\\
=&\frac{3}{\pi^2 r^3}  \int_{0}^{r} \D r' {r'}^2 \int k^2 \D k j_0(k r') \dv{\ln \G}{\ln a}\frac{\G^2(k,a)}{\G^2_0(k)} P^{\rm{lin}}_0(k) \mathcal{B}(k,a) \mathcal{N}(k,a)
\end{split}
\end{equation}
where we have additionally defined
\begin{equation}
\begin{split}
\mathcal{N}(k,a)=&\frac{1}{\bar{n}(a)}\int_{M_{\rm{min}}}^{M_{\rm{max}}} \D m\ n(m,a) \widetilde{W}[k R(m)].
\end{split}
\end{equation}
This factor arises from the integral over the window function without matching factor in the bias which was introduced when we took derivatives of the bias in Eq. (\ref{eq:correlation_derivative}).

The mean pairwise velocity then becomes
\begin{equation}\label{eq:scale_independent_velocities_windowed}
\begin{split}
V_h=\mean{v_{12}}_m=&- H r\frac{\meanM{\dv{\bar{\xi}_h}{\ln a}}_m}{3\left[1+ \mean{\xi_h}_m\right]}.
\end{split}
\end{equation}

If $\widetilde{W}(x)=1$ and $\G(k,a)=D(a)$, Eqs.~(\ref{eq:mass+volume_averaged_halo_correlation_derivative_app}) and (\ref{eq:scale_independent_velocities_windowed}) reduce to the expressions presented in Ref. \cite{Sheth:2000ff}. They do not, however, agree with the expressions presented in Refs.~ \cite{Bhattacharya:2007sk,Mueller:2014nsa}. In particular, the halo bias term differs between these two models. Instead of the term $\mathcal{B}^2(k,a)$ that appears in Eq.~(\ref{eq:mass_averaged_halo_correlation}), Refs. \cite{Bhattacharya:2007sk,Mueller:2014nsa} define
\begin{equation}\label{eq:lit_bias_term}
	b_h^{(q)}(k,a)=\frac{\int_{M_{\rm{min}}}^{M_{\rm{max}}} \D m\ m\ n(m,a) b^q(m,a) \widetilde{W}^2\left[k R(m)\right]}{\int_{M_{\rm{min}}}^{M_{\rm{max}}} \D m\ m\ n(m,a) \widetilde{W}^2\left[k R(m)\right]}.
\end{equation}
This is manifestly not equivalent to the expression above. Similarly, in Eq. (\ref{eq:mass+volume_averaged_halo_correlation_derivative_app}), our term $\mathcal{B}(k,a)\mathcal{N}(k,a)$ is replaced by $b_h^{(1)}(k,a)$. Equation~(\ref{eq:lit_bias_term}) does also not reduce to the bias expected in Ref. \cite{Sheth:2000ff} because setting $\widetilde{W}(x)=1$ does not yield $\xi_h = \bar{b}^2(a)\xi_{\rm{lin}}$ where $\bar{b}(a)$ would be the averaged halo bias as specified in Ref. \cite{Sheth:2000ff}. Instead, Eq. (\ref{eq:lit_bias_term}) leads to $$\xi_h =\frac{\int \D m\ m\ n(m, a) b^2(m,a)}{\int \D m\ m\ n(m, a)} \xi_{\rm{lin}}$$ i.e., the sample average (modulus some mass weighting) of the squared halo bias rather than the square of the averaged halo bias.  

\begin{figure}
	\includegraphics[width=0.5\linewidth]{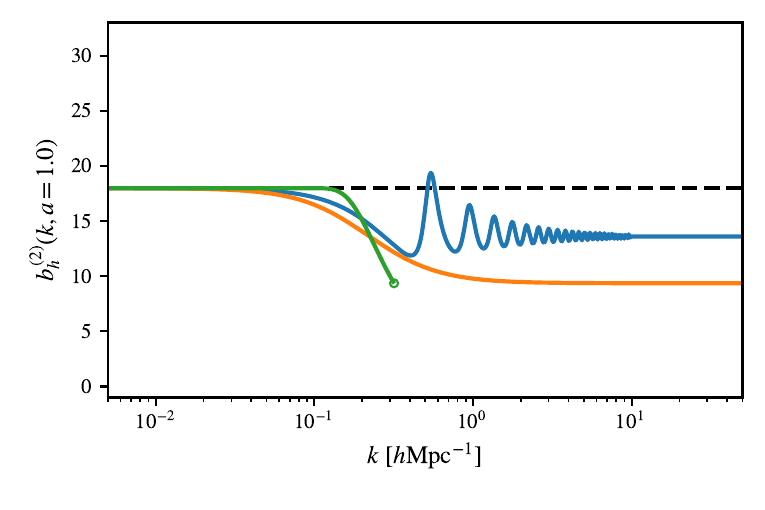}%
	\includegraphics[width=0.5\linewidth]{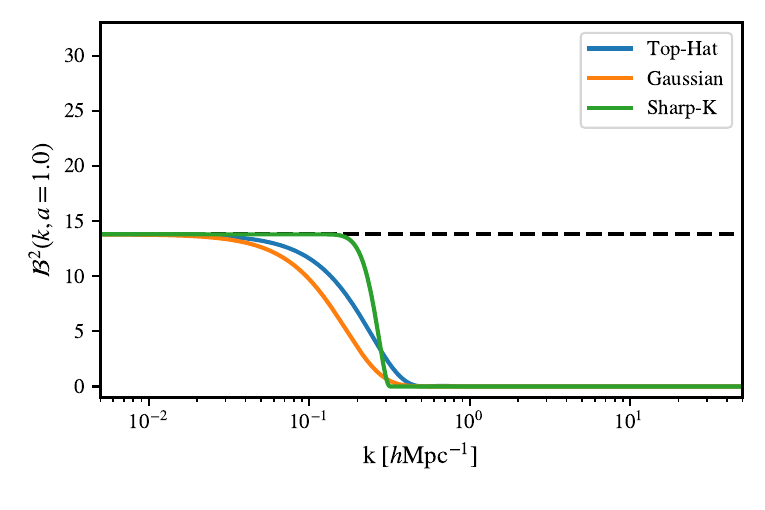}
	\caption{Comparing different bias prescriptions at $z=0.0$ for our fiducial $\Lambda$CDM model. \emph{Left}: Bias as presented by Refs.~\cite{Bhattacharya:2007sk,Mueller:2014nsa}. The dashed lines indicated the analytically computed asymptotic limit. \emph{Right}: Bias as computed using Eq.~(\ref{eq:mass_averaged_halo_bias_app}). \label{fig:bias_comp}}
\end{figure}

The bias term as given in Eq.~(\ref{eq:lit_bias_term}) exhibits some unexpected behavior at large $k$. When choosing a Top-Hat or Gaussian filter for $\widetilde{W}(x)$, the bias asymptotically approaches a finite, nonzero value at large $k$ (as seen in Fig.~\ref{fig:bias_comp}). That appears counterintuitive since it implies that the sample traces even scales smaller than $R(M_{\rm{min}})$. We would expect to see the bias approach zero for $k\gg1/R(M_{\rm{\min}})$. This problem does not arise with the bias expression from Eq. (\ref{eq:mass_averaged_halo_bias_app}). Furthermore, Eq.~(\ref{eq:lit_bias_term}) becomes undefined for large $k$ when using a sharp filter in $k$-space $\widetilde{W}(x) =1$ for $x\leq1$ and $0$ otherwise, as the denominator will evaluate to zero for $k>1/R(M_{\rm{min}})$ making the bias undefined. As discussed above, we use sharp-$k$ filters because they yield more accurate halo formation histories than other filters in structure suppressing models.  

\begin{figure}
	\includegraphics[width=0.5\linewidth]{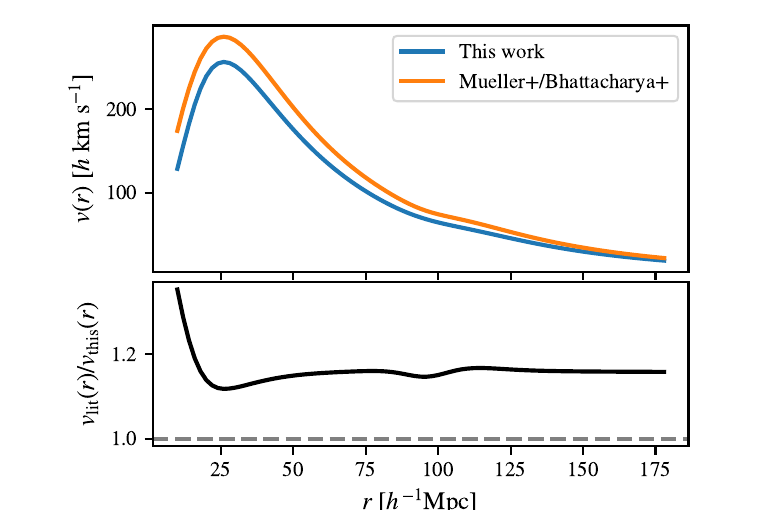}
	\caption{Comparing the mean pairwise velocity obtained with these two prescriptions for our fiducial $\Lambda$CDM model. As in Ref.~\cite{Mueller:2014nsa}, we adopt a Gaussian filter here. While the difference is large on small scales, it approaches a constant factor of $\sim1.15$ over the range of scales used in the analysis.\label{fig:comparing_mpv}}
\end{figure}

The impact of our modifications is shown in Fig.~\ref{fig:comparing_mpv}, where we adopt a Gaussian filter as in Ref. \cite{Mueller:2014nsa}. Not unexpectedly, the difference is largest at very small scales which are not usually used in the analysis because of observational uncertainties. Even on large scales, however, there remains an overall normalization difference.

Lastly, it remains to justify our approximation for Eq.~(\ref{eq:bias_deriv}),
\begin{equation}
	\dv{b}{\ln a} \simeq \dv{\ln \G}{\ln a}\left[1-b(m,a)\right].\label{apeq:bias_deriv}
\end{equation}
For this purpose, we can rewrite the $\ln a$ derivative of $b$ in terms of derivatives with respect to $\sigma_m(a)$ as
\begin{equation}
	\dv{b}{\ln a} = \dv{\sigma_m(a)}{\ln a} \dv{b}{\sigma_m(a)} = \dv{\ln \sigma_m(a)}{\ln a} \left[1-b(m,a)\right].
\end{equation}
We can compute $\dv{\ln \sigma_m(a)}{\ln a}$ as
\begin{equation}
    \dv{\ln \sigma_m(a)}{\ln a} = \frac{1}{2} \dv{\ln \sigma^2_m(a)}{\ln a} = \frac{1}{2 \pi^2 \sigma_m^2(a)} \int k^2 \D k \dv{\ln \G}{\ln a} \frac{\G^2(k,a)}{\G_0^2(k)} P^{\rm{lin}}_0(k) \widetilde{W}[k R(m)].\label{eq:apsigder}
\end{equation}
After mass averaging, if we use Eq.~(\ref{eq:apsigder}) instead of the right-hand side of Eq.~(\ref{apeq:bias_deriv}), we obtain the following instead 
 of the factor $\dv{\ln \G}{\ln a}\mathcal{N}(k,a)$ in Eq. (\ref{eq:mass+volume_averaged_halo_correlation_derivative_app}):
\begin{equation}
    \dv{\ln \G}{\ln a} \mathcal{B}(k,a) + \frac{1}{\bar{n}} \int \D m\  n(m)\dv{\ln \sigma_m(a)}{\ln a}\left[1-b(m,a)\right]\widetilde{W}\left[k R(m)\right].\label{eq:realfac}
\end{equation}
As mentioned in the body of the paper, if the scale dependence is weak, our approximation is exact. For small axion masses the axion abundance is strongly constrained and thus we expect only relatively weak scale dependence in the late time growth rate. For large axion masses on the other hand, while their abundance is relatively unconstrained they act increasingly like cold dark matter and introduce only weak scale dependence as well. We compare the numerical value for Eq.~(\ref{eq:realfac}) to $\dv{\ln \G}{\ln a}\mathcal{N}(k,a)$ and find that within the range of masses and abundances allowed at least by a SIV-like survey the difference is never larger than $\sim$4\% even for the most strongly scale-dependent cases allowed by our forecast ($m_a =10^{-27}$eV and $\eta_a=0.1$). This increases to about 20\% for axion masses of $m_a =10^{-27}$eV and $\eta_a=0.25$. For any masses larger than $m_a=10^{-26}$eV, the inaccuracies due to this approximation are at the subpercent level for all axion abundances. We thus expect the use of Eq.~(\ref{apeq:bias_deriv}) to induce deviations no worse than $1\%-20\%$ induced deviations in halo mass-function averaged predictions for $v(r)$. We reran our Fisher forecasts for a subset of our mass range (below $10^{-26}$eV) and found that our approximation has a negligible impact on the predicted detection limits ($\lesssim 4\%$).

\section{Numerical treatment of Ostriker-Vishniac integrals}\label{app:num}
\renewcommand\thefigure{\thesection.\arabic{figure}}    
\setcounter{figure}{0}    

We note that the integral to be evaluated to obtain $S(k)$ [Eq.~(\ref{eq:vish_spectrum})] appears singular at $x=y=1$. We argue here that this singularity behaves as $\epsilon^{-n}$ for $0<n<1$ and is thus integrable. For the purposes of this argument, we will assume that the growth function is approximately scale independent, i.e., $\G(k,a) \approx D(a)$, which is true on large scales. With this approximation the integrand becomes
\begin{equation}
I(x, y)=P(k y) P(k \sqrt{1+y^2-2 x y}) \frac{(1-x^2)(1-2xy)^2}{(1+y^2-2xy)^2}.
\end{equation}
The power spectrum $P(k)$ falls off quickly at large $k$ and so the contribution from such modes is small. At sufficiently small $k$, $P(k) \propto k^n$ where $n$ is the tilt of the power spectrum. Therefore, the integrand goes as
\begin{equation}
I(x,y) \propto k^{2n} y^n \frac{(1-x^2)(1-2xy)^2}{(1+y^2-2xy)^{2-0.5n}}.
\end{equation}
Expanding to first order around $x=y=1$, we find
\begin{equation}
\begin{split}
I(1-\epsilon,1+\delta) \propto&\frac{k^{2n}}{2^{2-0.5n}} \frac{(1+ n \delta) (2\epsilon) (1+ 4 \delta - 4\epsilon)}{\epsilon^{2-0.5n}}\\
\approx&\frac{k^{2n}}{2^{2-0.5n}} \frac{2\epsilon}{\epsilon^{2-0.5n}}\\
=&\frac{k^{2n}}{2^{1-0.5n}} \frac{1}{\epsilon^{1-0.5n}}.
\end{split}
\end{equation}
In the expression above, we have made use of the fact that $n$ is observationally constrained to be close to unity. We can now see that for any physically reasonable value of $n$ the singularity should be integrable.

In order to numerically evaluate the integral, we perform a coordinate transform $x\to t$. Since the singularity has the form $1/(1-x)^{1-0.5n}$, one can require $\dv{x}{t} \propto(1-x)^{1-0.5n}$. If we then redefine the integrand in terms of $t$, we will have multiplied out the divergent factor. This implies (up to scalar factors) $t=(1-x)^{n/2}$. With this transformation,
\begin{equation}
\int_{-1}^{1} I(x, y) \D x = {2\over n} \int_{0}^{2^{n/2}} t^{2/n-1} I(1-t^{2/n}, y) \D t .
\end{equation}

\section{Mean pairwise velocity parameter degeneracies}
\label{sec:degen}
\renewcommand\thefigure{\thesection.\arabic{figure}}    
\setcounter{figure}{0}    

\begin{figure}
	\includegraphics[width=\linewidth]{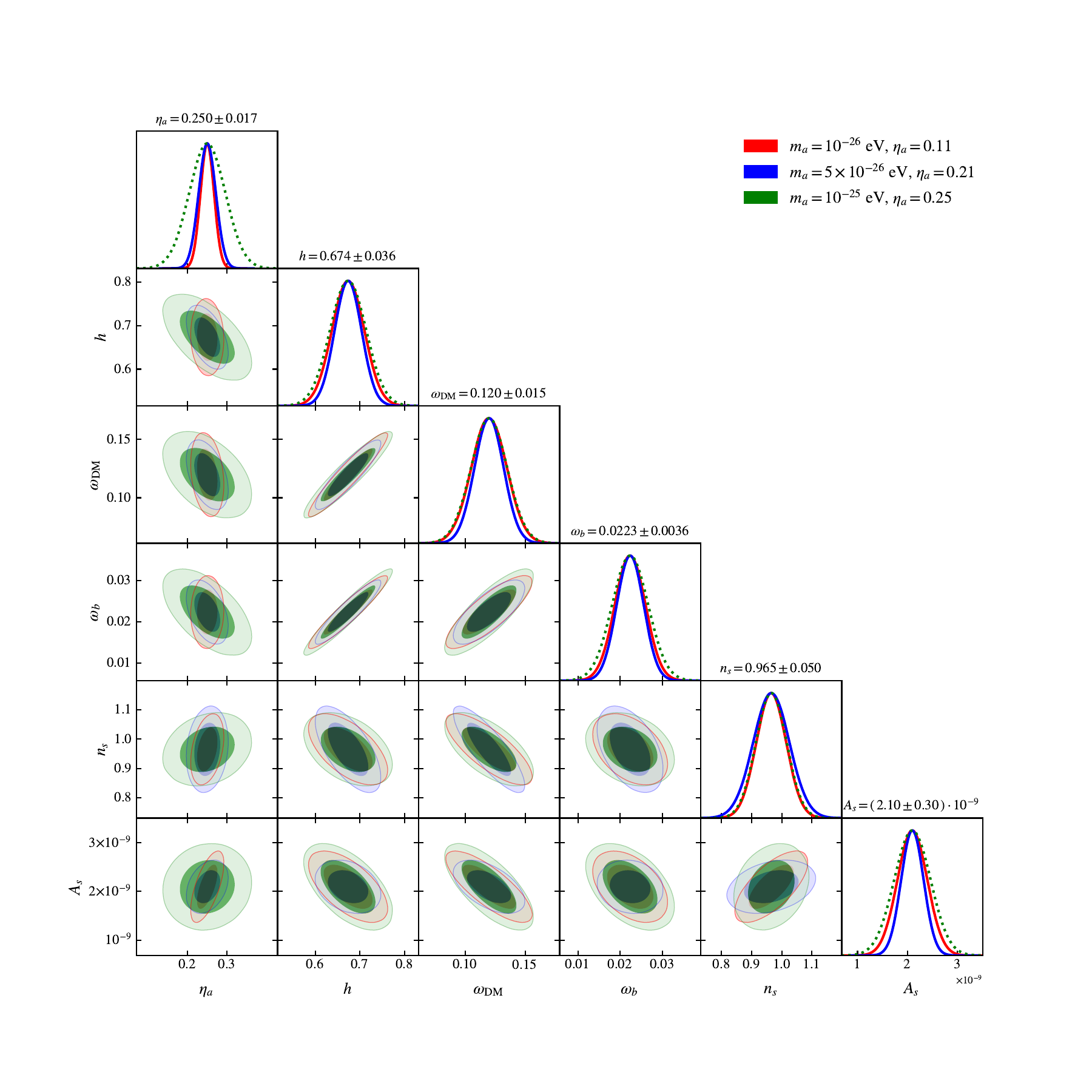}%
	\caption{Parameter degeneracy forecasts from our Fisher-matrix analysis. This figure shows degeneracies between the ULA fraction $\eta_{a}$ and the standard $\Lambda$CDM cosmological parameters. The symbol $\omega_{i}=\Omega_{i}h^{2}$ for species $i \in \left\{{\rm DM}, b\right\}$.}
	\label{fig:tri1}
\end{figure}

\begin{figure}
		\includegraphics[width=\linewidth]{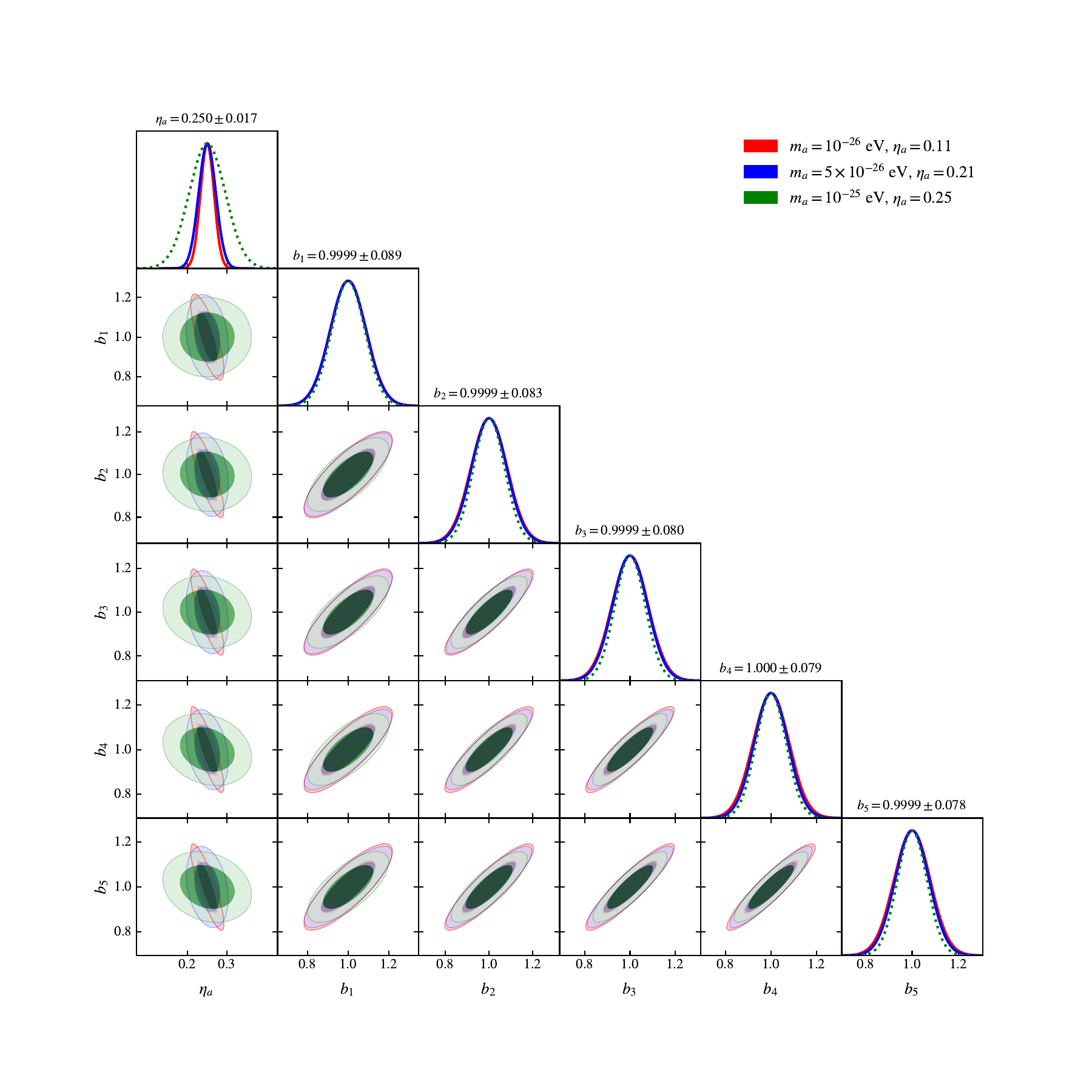}
	\caption{Parameter degeneracy forecasts from our Fisher-matrix analysis. This figure shows degeneracies between the bias nuisance parameters $b_{1}$, $b_{2}$, $b_{3}$, $b_{4}$, and $b_{5}$ as well as the ULA fraction $\eta_a$.}
		\label{fig:tri2}

\end{figure}

\begin{figure}
		\includegraphics[width=\linewidth]{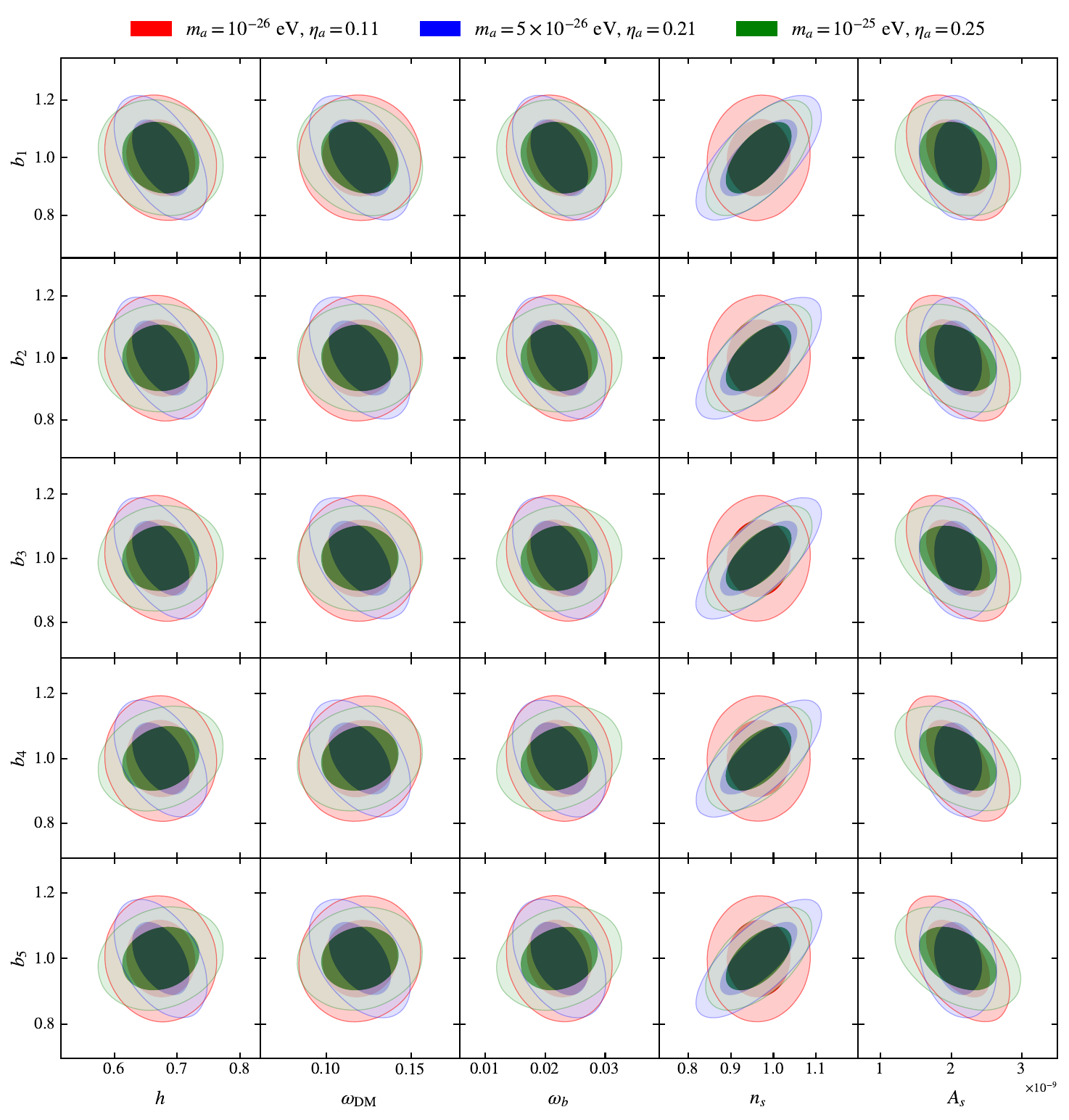}%
	\caption{Parameter degeneracy forecasts from our Fisher-matrix analysis. This figure shows degeneracies between the bias nuisance parameters $b_{1}$, $b_{2}$, $b_{3}$, $b_{4}$, and $b_{5}$ with the standard cosmological parameters. The symbol $\omega_{i}=\Omega_{i}h^{2}$ for species $i \in \left\{{\rm DM}, b\right\}$.}
		\label{fig:tri3}

\end{figure}

\begin{figure}
\includegraphics[width=\linewidth]{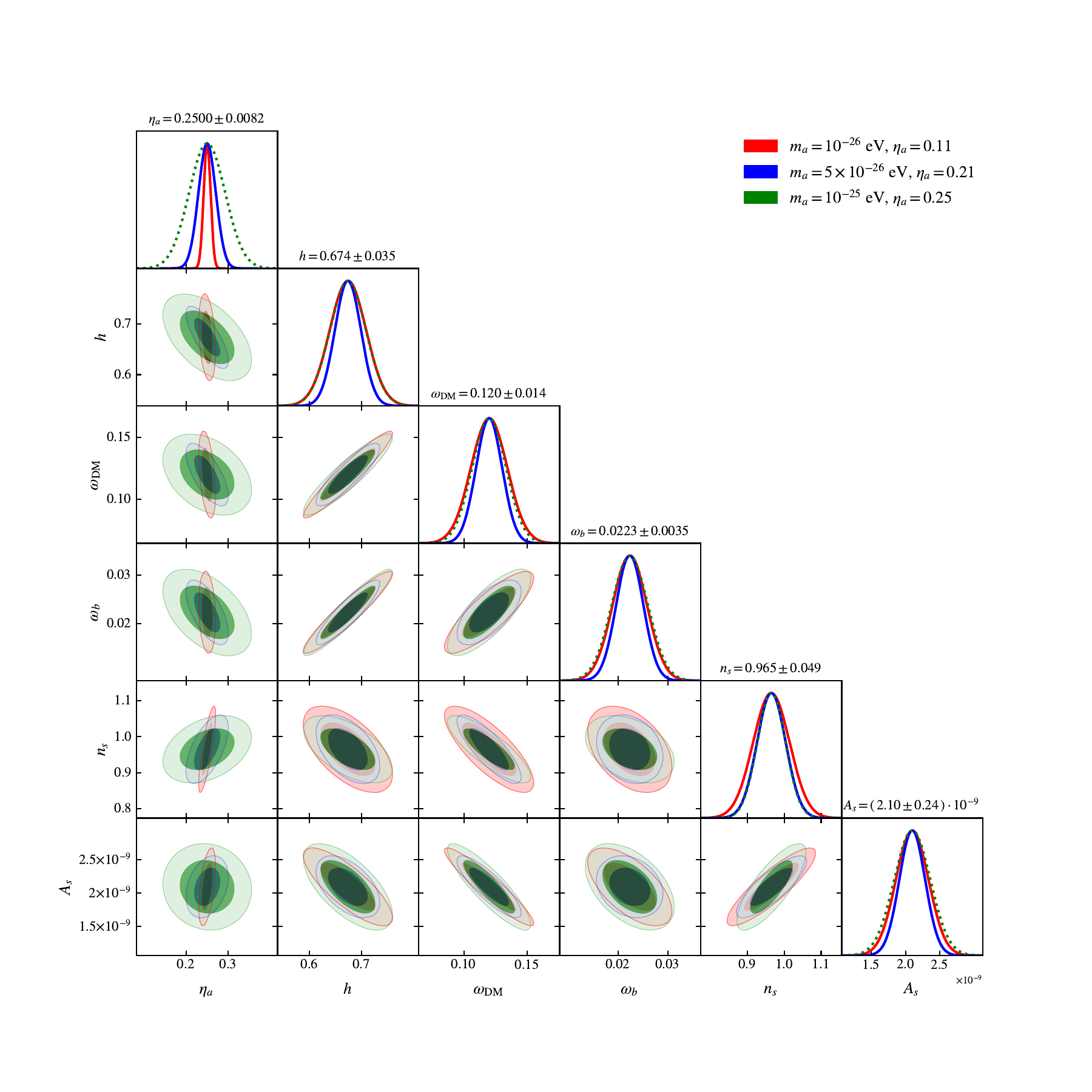}
	\caption{Parameter degeneracy forecasts from our Fisher-matrix analysis. This figure shows degeneracies between the ULA abundance $\eta_{a}$ and $\Lambda$CDM parameters when bias nuisance parameters are not marginalized over. The symbol $\omega_{i}=\Omega_{i}h^{2}$ for species $i \in \left\{{\rm DM}, b\right\}$.}
	\label{fig:tri4}

\end{figure}

In order to inspect degeneracies between different parameters in our analysis, we draw $10^6$ random samples from a multidimensional Gaussian distribution with covariance given by the inverse of the Fisher matrix computed as described in Eq. (\ref{eq:fisher_matrix_expression}). The samples drawn are then analyzed using \textsc{GetDist}.\footnote{https://github.com/cmbant/getdist} Degeneracies between the $\Lambda$CDM cosmological parameters and the ULA fraction, the bias parameters and the ULA fraction, and the $\Lambda$CDM cosmological parameters and the bias parameters are shown in Figs.~\ref{fig:tri1}, \ref{fig:tri2}, and \ref{fig:tri3} respectively. We also show the degeneracies obtained when neglecting the marginalization over the bias parameters (Fig.~\ref{fig:tri4}).  


\twocolumngrid

\bibliography{ksz}

\end{document}